# Perturbative nonlinear J-matrix method of scattering in two dimensions

T. J. Taiwo[(a)], A. D. Alhaidari[(b)] and U. Al Khawaja[(c,a)]

[(a)] *Physics Department, Untied Arab Emirate University, P.O. Box 15551, Al-Ain, United Arab Emirates*
[(b)] *Saudi Center for Theoretical Physics, P. O. Box 32741, Jeddah 21438, Saudi Arabia*
[(c)] *Department of Physics, School of Sciences, University of Jordan, Amman 11942, Jordan*

**Abstract:** We introduce a perturbative formulation for a nonlinear extension of the J-matrix method of scattering in two dimensions. That is, we obtain the scattering matrix for the time-independent nonlinear Schrödinger equation in two dimensions with circular symmetry. The formulation relies on the linearization of products of orthogonal polynomials and on the utilization of the tools of the J-matrix method. Gauss quadrature integral approximation is instrumental in the numerical implementation of the approach. We present the theory for a general $\psi^{2n+1}$ nonlinearity, where *n* is a natural number, and obtain results for the cubic and quintic nonlinearities, $\psi^3$ and $\psi^5$. At certain value(s) of the energy, we observe the occurrence of bifurcation with two stable solutions. This curious and interesting phenomenon is a clear signature and manifestation of the underlying nonlinearity.

## 1. Introduction

Nonlinear Physics is fundamental for understanding real-world phenomena in, for instance, nonlinear optics [1], turbulent fluid dynamics [2], plasma physics [3], chaotic systems [4], Bose-Einstein condensates [5], biological systems [6], and many others [7]. Novel phenomena emerge due to the nonlinearity in these systems, such as solitons, chaos, fractals, and phase transitions with important applications, such as solitons in optical fibers and waveguide arrays [1]. Mathematically, nonlinear systems are described by nonlinear differential equations, either in continuous or in discrete form [5]. In particular, nonlinear phenomena are typically associated to solutions of these equations, hence the interest in the existence and properties of the solutions. A plethora of methods have been developed to find exact solutions to nonlinear differential equations, investigate their integrability, and study the stability of their solutions [8]. Nonlinear differential equations have fundamentally different features in comparison with their linear counterparts. For example, the number of fundamentally different solutions is not limited by the largest derivative, as in linear differential equations. In principle, the number of independent solutions can be infinite, in fact it can be families with infinite number of members in each family. Another important difference is the inapplicability of the superposition principle; the linear combination of two independent solutions is not generally a third solution, except for some very specific cases [9].

The nonlinear Schrödinger equation (NLSE) is one of the most important nonlinear evolution equations as it described many major fields in physics including matter-waves of Bose-Einstein condensates [10], nonlinear optics [1], and ocean waves [11]. Its solitonic solutions, particularly bright and dark solitons, exhibit the unique feature of preserving their integrity after mutual interactions, or even after scattering by reflectionless potentials. The integrability of the so-called fundamental NLSE in one dimension was established through the Lax pair and Inverse scattering Transform method by Zakharov and Shabat [12], and a large number of its exact solutions is known [13]. However, many versions of the NLSE – sometimes called the nonautonomous NLSE – are not integrable. In such cases, one resorts to numerical solutions. It should be noted here that while a certain version of the NLSE may be nonintegrable, it may nonetheless have some exact analytical solutions, let alone numerical solutions. A striking example is the NLSE in two- and three-dimensions. For instance, the NLSE in two dimensions

$$i\frac{\partial}{\partial t}\Phi(r,\theta,t) = -\frac{1}{2}\left(\frac{\partial^2}{\partial r^2} + \frac{1}{r}\frac{\partial}{\partial r} + \frac{1}{r^2}\frac{\partial^2}{\partial \theta^2}\right)\Phi(r,\theta,t) + g|\Phi(r,\theta,t)|^2\Phi(r,\theta,t) \quad (1)$$

is generally not integrable, but admits important stable numerical solutions characterized by a number of nodes. Here, $\Phi(r,\theta,t)$ is a complex field, $r$ and $\theta$ are the cylindrical coordinates, and $g$ is a real coupling parameter. For stationary solutions of the form

$$\Phi(r,\theta,t) = \Psi(r)e^{-i(\lambda t + \alpha\theta)}, \quad (2)$$

the NLSE equation takes the time-independent form

$$-\frac{1}{2}\left(\Psi'' + \frac{1}{r}\Psi' - \frac{\alpha^2}{r^2}\Psi\right) + g\Psi^3 = \lambda\Psi, \quad (3)$$

where $\lambda$ and $\alpha$ are real constants and $\Psi(r)$ is a real function, and the prime denotes a derivative with respect to $r$. Solving this equation numerically shows that it supports a family of infinite number of localized nodeless and nodal solutions [13]. A similar situation occurs for the NLSE with external potentials. Numerical solutions show again that there are localized solutions which are confined by the potential and exist in stable nodeless and unstable nodal forms [14].

Scaling transformation may be used to show that

$$\Phi(r,\theta,t) = r^{(1-D)/2}e^{i\sqrt{-(D-1)(D-3)}\,\theta/2}\Psi(r,t)e^{-i(\lambda t + \alpha\theta)}, \quad (4)$$

is a solution to the following NLSE

$$i\frac{\partial}{\partial t}\Phi(r,\theta,t) = -\frac{1}{2}\left(\frac{\partial^2}{\partial r^2} + \frac{D-1}{r}\frac{\partial}{\partial r} + \frac{1}{r^2}\frac{\partial^2}{\partial \theta^2}\right)\Phi(r,\theta,t) + gr^{D-1}|\Phi(r,\theta,t)|^2\Phi(r,\theta,t), \quad (5)$$

provided that $\Psi(r,t)$ is a solution to the fundamental NLSE. Here $D = 1,2,3$, corresponds to the one-, two-, and three-dimensional cases. The operator between brackets corresponds to the Laplacian in $D$ dimensions. This transformation provides a map that links solutions of the fundamental NLSE with those of the above version of NLSE in $D$ dimensions. Due to the factor $r^{(1-D)/2}$ in the above function transformation, the solution will be diverging at $r = 0$ for $D = 2$, and $D = 3$. The divergence is removed when the solution $\Psi(r,t)$ of the fundamental NLSE goes to zero at $r = 0$ as, or faster than, $r$. This is the case for example with $\Psi(r,t)$ being the dark soliton solution, which is given in terms of the hyperbolic tangent function.

In this study, we focus on the scattering aspect of nonlinearity and employ the tools of the J-matrix method to obtain the scattering matrix. We limit our treatment to a single-channel elastic scattering. Our choice of the J-matrix stems from our long experience with the method and its success in a wide range of applications in atomic, molecular, and nuclear physics [15]. Additionally, it is the only algebraic scattering method that relies heavily on the analytic power of orthogonal polynomials, which turns out to be an essential ingredient in our linearization technique towards finding a solution. However, the method was developed long ago to handle only linear interactions with short range potentials. Therefore, it has first to be extended to the nonlinear domain. The first attempt to do that was carried out by the first two authors but using a toy model [16]. Here, we go further and eliminate some of the limitations in that toy model. For example, we go beyond the toy model and do not use an arbitrary choice of ansatz for the solution. Moreover, we include a linear potential in addition to the nonlinear self-interaction. Furthermore, we present the theory for a general $\psi^{2n+1}$ nonlinearity in the NLSE, where $n = 1, 2, 3,...$. However, we were only able to advance a perturbative solution to the problem which limits its validity to weak coupling.

Readers are strongly encouraged to go through our first attempt at a nonlinear extension of the J-matrix method in [16]. For an expanded awareness of the J-matrix method of scattering, one may refer to any of the vast literature on the subject such as those listed in [15, 17-20].

## 2. Preliminaries

We consider the following NLSE in the presence of a linear potential $V(\vec{r})$ that reads (in the atomic units $\hbar = M = 1$):

$$i\frac{\partial}{\partial t}\Phi(\vec{r},t) = -\frac{1}{2}\vec{\nabla}^2\Phi(\vec{r},t) + V(\vec{r})\Phi(\vec{r},t) + g|\Phi(\vec{r},t)|^{2n}\Phi(\vec{r},t), \qquad (6)$$

where $n$ is a natural number, $n = 1, 2, 3,..$, and the coupling parameter $g$ is dimensionless if the conformal degree of the wavefunction $\Phi(\vec{r},t)$ is $\mathcal{C} = 1/n$. In three dimensions, $\vec{\nabla}^2 = \frac{\partial^2}{\partial r^2} + \frac{2}{r}\frac{\partial}{\partial r} + \frac{1}{r^2}L(\theta,\varphi)$ and $L(\theta,\varphi) = \frac{\partial^2}{\partial \theta^2} + \frac{\cos\theta}{\sin\theta}\frac{\partial}{\partial \theta} + \frac{1}{\sin^2\theta}\frac{\partial^2}{\partial \varphi^2}$. For a spherically symmetric potential $V(\vec{r}) = V(r)$, the stationary solution is written as follows

$$\Phi(\vec{r},t) = \frac{1}{r}\psi(r)Y(\theta,\varphi)e^{-iEt}, \qquad (7)$$

where $Y(\theta,\varphi) = \sqrt{\frac{2\ell+1}{4\pi}\frac{(\ell-\mathsf{m})!}{(\ell+\mathsf{m})!}}\, e^{i\mathsf{m}(\varphi+\pi)}P_\ell^\mathsf{m}(\cos\theta)$ and $P_\ell^\mathsf{m}(\cos\theta)$ is the Legendre polynomial with $\ell$ being the orbital angular momentum quantum number, $\mathsf{m} = -\ell, -\ell+1,.., \ell$ and $E$ is the energy. Moreover, $L(\theta,\varphi)Y(\theta,\varphi) = -\ell(\ell+1)Y(\theta,\varphi)$. Substituting in (6), we obtain

$$E\psi(r) = \left[-\frac{1}{2}\frac{d^2}{dr^2} + \frac{\ell(\ell+1)}{2r^2} + V(r)\right]\psi(r) + g\,r^{-2n}[y(\theta)]^{2n}|\psi(r)|^{2n}\psi(r). \qquad (8)$$

where $y(\theta) = \sqrt{\frac{2\ell+1}{4\pi}\frac{(\ell-\mathsf{m})!}{(\ell+\mathsf{m})!}}\, P_\ell^\mathsf{m}(\cos\theta)$ and $g$ is of length dimension $2(n-1)$ (so the canonical linearity is cubic corresponding to $n = 1$). Consequently, the NLSE (6) is not separable in 3D and will not be pursued in this work. On the other hand, in two dimensions with $\vec{\nabla}^2 = \frac{\partial^2}{\partial r^2} + \frac{1}{r}\frac{\partial}{\partial r} + \frac{1}{r^2}\frac{\partial^2}{\partial \varphi^2}$, the stationary solution with circular symmetry is written as follows

$$\Phi(\vec{r},t) = \frac{1}{\sqrt{r}}\psi(r)Y(\varphi)e^{-iEt}, \qquad (9)$$

where $Y(\varphi) = e^{i m \varphi}/\sqrt{2\pi}$ and $\mathsf{m}$ is the azimuthal quantum number $\mathsf{m} = 0, \pm 1, \pm 2, \ldots$ Substituting in (6), we obtain the following nonlinear radial Schrödinger equation

$$\left[-\frac{1}{2}\frac{d^2}{dr^2} + \frac{\mathsf{m}^2 - \frac{1}{4}}{2r^2} + V(r) - E\right]\psi(r) = -g\, r^{-n} |\psi(r)|^{2n} \psi(r), \tag{10}$$

where $g$ is of length dimension $n-2$ (so the canonical linearity is quintic with $n = 2$). Therefore, the radial component of the wavefunction is fully separable and our solution could become useful in condensed matter for 2D materials. The linear potential $V(r)$ can be analytic or nonanalytic function but is assumed to be non-singular and short-range. The solution of Eq. (10) is not always stable (i.e., it may not exist) for arbitrary values of the physical parameters. However, for a fixed set of values of the parameters $\{n, g, \mathsf{m}\}$ and potential $V(r)$, the solution could become stable for either a continuous and infinite range of $E$, a number of continuous and finite but disconnected intervals of $E$, a set of discrete values of $E$, or a combination thereof. Assuming weak nonlinear coupling, our solution will be perturbative where Eq. (10) is solved at each order of the perturbation series as follows:

$$\mathcal{D}\psi_0(r) = 0, \quad \mathcal{D}\psi_1(r) = -g\, r^{-n} |\psi_0(r)|^{2n} \psi_1(r), \quad \mathcal{D}\psi_2(r) = -g\, r^{-n} |\psi_1(r)|^{2n} \psi_2(r), \ldots \tag{11}$$

$$\mathcal{D}\psi_m(r) = -g\, r^{-n} |\psi_{m-1}(r)|^{2n} \psi_m(r). \tag{12}$$

where $\mathcal{D} = -\frac{1}{2}\frac{d^2}{dr^2} + \frac{\ell^2 - \frac{1}{4}}{2r^2} + V(r) - E$, and $\ell = |\mathsf{m}|$. Assuming convergence and full knowledge of the linear solution $\psi_0(r)$, the exact solution is $\psi(r) = \lim_{m \to \infty}[\psi_m(r)]$ where $\|\psi_{m+1}(r) - \psi_m(r)\| \to 0$. Consequently, for weak coupling $\psi_m(r)$ could become a good approximation for small enough perturbation order $m$.

## 3. J-matrix formulation of the problem

Let $\{\phi_k(r)\}_{k=0}^{\infty}$ be a complete set of *real* and square integrable functions that forms an invariant subspace in the domain of the free wave operator $\mathcal{D}_0 = -\frac{1}{2}\frac{d^2}{dr^2} + \frac{\ell^2 - \frac{1}{4}}{2r^2} - E$ (i.e., $\mathcal{D}_0 \phi_k \in \{\phi_l\}_{l=0}^{\infty}$). Specifically, we require that

$$\mathcal{D}_0 \phi_k(r) = \omega(r)\left[(a_k - z)\phi_k(r) + b_{k-1}\phi_{k-1}(r) + b_k \phi_{k+1}(r)\right], \tag{13}$$

where $z$ is an energy dependent parameter (called the *spectral parameter*) and $\{a_k, b_k\}$ are real constants that are independent of $z$ such that $b_k \neq 0$ for all $k$. $\omega(r)$ is an entire function that does not vanish (except possibly at the boundaries $r = 0$ and $r \to \infty$). Therefore, we can always expand the wavefunction in this complete basis set as $\psi_m(r) = \sum_{k=0}^{\infty} A_k^{(m)}(E)\phi_k(r)$ where $\{A_k^{(m)}(E)\}_{k=0}^{\infty}$ are *complex* expansion coefficients that contain all the physical properties of the system at the perturbative order $m$ whereas the basis elements $\{\phi_k(r)\}$ are "physically dummy".

For the problem described by Eq. (10), we take the following orthonormal basis elements (sometimes, known as the "oscillator basis"):

$$\phi_k(r) = \sqrt{\frac{2\lambda\Gamma(k+1)}{\Gamma(k+\nu+1)}} (\lambda r)^\alpha e^{-\lambda^2 r^2/2} L_k^\nu(\lambda^2 r^2) = \sqrt{\frac{2\lambda}{\Gamma(\nu+1)}} (\lambda r)^\alpha e^{-\lambda^2 r^2/2} \tilde{L}_k^\nu(\lambda^2 r^2), \quad (14)$$

where $L_k^\nu(x)$ is the conventional Laguerre polynomial and $\tilde{L}_k^\nu(x)$ is its normalized version, $\tilde{L}_k^\nu(x) = \sqrt{\frac{\Gamma(k+1)\Gamma(\nu+1)}{\Gamma(k+\nu+1)}} L_k^\nu(x) = \sqrt{\frac{\Gamma(k+\nu+1)}{\Gamma(k+1)\Gamma(\nu+1)}} {}_1F_1(-k;\nu+1;x)$. $\lambda$ is a non-physical real positive scale parameter of inverse length dimension. Using the differential equation of the Laguerre polynomials, their differential property and recursion relation, we can show that these basis elements satisfy the requirement (13) with $\alpha = \ell + \frac{1}{2}$, $\nu = \ell$, $\omega(r) = 1$, $z = E$, and

$$a_k = \frac{\lambda^2}{2}(2k+\ell+1), \quad b_k = \frac{\lambda^2}{2}\sqrt{(k+1)(k+\ell+1)}, \quad (15)$$

Therefore, the left-hand side of Eq. (12) reads

$$\sum_{k=0}^\infty \phi_k(r)\left[(a_k - E)A_k^{(m)}(E) + b_{k-1}A_{k-1}^{(m)}(E) + b_k A_{k+1}^{(m)}(E)\right] + V(r)\sum_{k=0}^\infty \phi_k(r) A_k^{(m)}(E), \quad (16)$$

whereas the right-hand side becomes

$$-g r^{-n} |\psi_{m-1}(r)|^{2n} \psi_m(r) = -g(2/\ell!)^{n+\frac{1}{2}} \lambda^{2n+\frac{1}{2}} (\lambda r)^{\ell(2n+1)+\frac{1}{2}} e^{-(2n+1)\lambda^2 r^2/2} \times$$
$$\sum_{k_1,k_2,...,k_{2n+1}=0}^\infty \left(A_{k_1}^{(m-1)} A_{k_2}^{(m-1)} ... A_{k_n}^{(m-1)}\right) \overline{\left(A_{k_{n+1}}^{(m-1)} A_{k_{n+2}}^{(m-1)} ... A_{k_{2n}}^{(m-1)}\right)} \tilde{L}_{k_1}^\ell(\lambda^2 r^2) \tilde{L}_{k_2}^\ell(\lambda^2 r^2) ... \tilde{L}_{k_{2n}}^\ell(\lambda^2 r^2) \tilde{L}_{k_{2n+1}}^\ell(\lambda^2 r^2) A_{k_{2n+1}}^{(m)} \quad (17)$$

In this equation, the product of the Laguerre polynomials in the sum can be linearized using the linearization of products of orthogonal polynomials [21,22] which gives

$$\tilde{L}_{k_1}^\ell(\lambda^2 r^2) \tilde{L}_{k_2}^\ell(\lambda^2 r^2) ... \tilde{L}_{k_{2n+1}}^\ell(\lambda^2 r^2) = \sum_{j=0}^{k_1+k_2+...+k_{2n+1}} C_{k_1,k_2,...,k_{2n+1}}^j \tilde{L}_j^\ell(\lambda^2 r^2), \quad (18)$$

where the linearization coefficients $C_{k_1,k_2,...,k_{2n+1}}^j$ could be obtained using the orthogonality of the Laguerre polynomials as

$$C_{k_1,k_2,...,k_{2n+1}}^j = \int_0^\infty \left(x^\ell e^{-x}/\ell!\right) \tilde{L}_j^\ell(x) \tilde{L}_{k_1}^\ell(x) \tilde{L}_{k_2}^\ell(x) ... \tilde{L}_{k_{2n+1}}^\ell(x) dx. \quad (19)$$

This integral could be evaluated using any robust numerical integration scheme such as the Gauss quadrature integral approximation, which goes as follows (see, for example, Ref. [23]): Let $J$ be the Jacobi matrix associated with the three-term recursion relation of the normalized Laguerre polynomials. That is, the tridiagonal symmetric matrix whose elements are

$$J_{i,j} = (2/\lambda^2)\left(a_i \delta_{i,j} - b_{i-1}\delta_{i-1,j} - b_i \delta_{i+1,j}\right). \quad (20)$$

Now, let $\{\xi_l\}_{l=0}^{N-1}$ be the real distinct eigenvalues of an $N \times N$ truncated version of $J$ and let $\{\Lambda_{i,l}\}_{i=0}^{N-1}$ be the corresponding normalized eigenvectors. Then, for a square integrable function $f(x)$ and for some large enough integer $N$, we can write

$$\frac{1}{\ell!}\int_0^\infty f(x)\left[x^\ell e^{-x}\tilde{L}_i^\ell(x)\tilde{L}_j^\ell(x)\right]dx \cong \sum_{l=0}^{N-1}\Lambda_{i,l}F_{l,l}\Lambda_{j,l} = \left(\Lambda F \Lambda^T\right)_{i,j}, \tag{21}$$

where $F$ is an $N \times N$ diagonal matrix with the elements $F_{i,j} = f(\xi_i)\delta_{i,j}$. Consequently, we obtain a good approximation for the integral (19) that reads $C^j_{k_1,k_2,...,k_{2n+1}} \cong \left(\Lambda U^{(\ell)}_{k_1,k_2,...,k_{2n}}\Lambda^T\right)_{j,k_{2n+1}}$ where, for a given set of indices $\{k_1,k_2,...,k_{2n}\}$, $U^{(\ell)}_{k_1,k_2,...,k_{2n}}$ is an $N \times N$ diagonal matrix with elements

$$\left(U^{(\ell)}_{k_1,k_2,...,k_{2n}}\right)_{i,j} = \tilde{L}^\ell_{k_1}(\xi_i)\tilde{L}^\ell_{k_2}(\xi_i)...\tilde{L}^\ell_{k_{2n}}(\xi_i)\delta_{i,j} = \left[\Lambda_{k_1,i}\Lambda_{k_2,i}...\Lambda_{k_{2n},i}\big/(\Lambda_{0,i})^{2n}\right]\delta_{i,j}, \tag{22}$$

and where we have used the formula $\tilde{L}^\ell_k(\xi_i) = \Lambda_{k,i}/\Lambda_{0,i}$. On the other hand, Gauss quadrature exact rule (see page 3 of Ref. [23]) states that the value of the integral $\int_0^\infty \rho(x)p_k(x)dx$ becomes exact and is equal to $\sum_{l=0}^{N-1}\Lambda_{0,l}^2 p_k(\xi_l)$ if $p_k(x)$ is a polynomial in $x$ of degree $k \leq 2N-1$, where the quadrature matrix $J$ is associated with the polynomial whose weight function is $\rho(x)$. Therefore, the integral (19) results in an exact evaluation as $C^j_{k_1,k_2,...,k_{2n+1}} = \left(\Lambda U^{(\ell)}_{k_1,k_2,...,k_{2n}}\Lambda^T\right)_{j,k_{2n+1}}$ if $2N-1$ is greater than or equal to $(j+k_1+k_2+...+k_{2n}+k_{2n+1})$. Nonetheless, using Theorem 1 in Section 2 of reference [21], we can also write

$$C^j_{k_1,k_2,...,k_{2n+1}} = \left[\tilde{L}^\ell_{k_1}(J)\tilde{L}^\ell_{k_2}(J)...\tilde{L}^\ell_{k_{2n}}(J)\right]_{j,k_{2n+1}} \tag{23}$$

It should be obvious from Eq. (19) that $C^j_{k_1,k_2,...,k_{2n+1}}$ is totally symmetric under the exchange of any two of its indices. However, this might not be evident from the definition (23) for a finite matrix $J$ in which case one needs to employ some numerical routines to insure this full symmetry. Moreover, since the left-hand side of Eq. (18) is a polynomial in $\lambda^2 r^2$ with a maximum degree of $k_1+k_2+...+k_{2n}+k_{2n+1}$ then so too is the right-hand side. Hence, for a given set of indices $\{k_i\}_{i=1}^{2n+1}$, $C^j_{k_1,k_2,...,k_{2n+1}} = 0$ if $j > k_1+k_2+...+k_{2n}+k_{2n+1}$ and that is why the sum in (18) terminates there. Thus, the exchange symmetry dictates that $C^j_{k_1,k_2,...,k_{2n+1}} = 0$ if any of its indices is greater than the sum of all other indices. That is for any integer $r \in \{1,2,..,2n+2\}$, $C^{k_{2n+2}}_{k_1,k_2,...,k_{2n+1}} = 0$ if $k_r > \sum_{i=1,i\neq r}^{2n+2} k_i$. In the technical notes of Appendix A, we address some numerical issues relevant to the calculation of $C^j_{k_1,k_2,...,k_{2n+1}}$ using Eq. (23) especially its total symmetry and the minimum size of the matrix $J$. Consequently, the right-hand side of Eq. (12) reads

$$\begin{aligned}&-g\left(2\lambda^2/\ell!\right)^n(\lambda r)^{2n\ell}e^{-n\lambda^2 r^2} \times \\ &\sum_{j,k_1,k_2,...,k_{2n+1}=0}^\infty \left(A^{(m-1)}_{k_1}A^{(m-1)}_{k_2}...A^{(m-1)}_{k_n}\right)\left(\overline{A^{(m-1)}_{k_{n+1}}A^{(m-1)}_{k_{n+2}}...A^{(m-1)}_{k_{2n}}}\right)A^{(m)}_{k_{2n+1}}C^j_{k_1,k_2,...,k_{2n+1}}\phi_j(r)\end{aligned} \tag{24}$$

Equating (16) to (24) and multiplying both sided by $\phi_i(r)$ then integrating over $r$, we obtain the following equation

$$E A_i^{(m)} = a_i A_i^{(m)} + b_{i-1} A_{i-1}^{(m)} + b_i A_{i+1}^{(m)} + \sum_{j=0}^{\infty} A_j^{(m)} \int_0^{\infty} \phi_i(r) V(r) \phi_j(r) dr + g\left(\lambda\sqrt{2/\ell!}\right)^{2n} \times$$

$$\sum_{j,k_1,k_2,\ldots,k_{2n+1}=0}^{\infty} \left(A_{k_1}^{(m-1)} A_{k_2}^{(m-1)} \ldots A_{k_n}^{(m-1)}\right) \left(\overline{A_{k_{n+1}}^{(m-1)} A_{k_{n+2}}^{(m-1)} \ldots A_{k_{2n}}^{(m-1)}}\right) A_{k_{2n+1}}^{(m)} C_{k_1,k_2,\ldots,k_{2n+1}}^j \int_0^{\infty} (\lambda r)^{2n\ell} e^{-n\lambda^2 r^2} \phi_i(r) \phi_j(r) dr \quad (25)$$

where we have used the orthogonality of the Laguerre polynomials. Defining $x := \lambda^2 r^2$ and $f^{(n,\ell)}(x) := x^{n\ell} e^{-nx}$, the two integrals in (25) read as follows

$$\int_0^{\infty} \phi_i(r) V(r) \phi_j(r) dr = \frac{1}{\ell!} \int_0^{\infty} V\left(\sqrt{x}/\lambda\right) \left[x^{\ell} e^{-x} \tilde{L}_i^{\ell}(x) \tilde{L}_j^{\ell}(x)\right] dx, \quad (26a)$$

$$\int_0^{\infty} (\lambda r)^{2n\ell} e^{-n\lambda^2 r^2} \phi_i(r) \phi_j(r) dr = \frac{1}{\ell!} \int_0^{\infty} f^{(n,\ell)}(x) \left[x^{\ell} e^{-x} \tilde{L}_i^{\ell}(x) \tilde{L}_j^{\ell}(x)\right] dx. \quad (26b)$$

The integral (26b) can be written as:

$$F_{i,j}^{(n,\ell)} := \sqrt{\frac{(i!)(j!)}{(i+\ell)!(j+\ell)!}} \int_0^{\infty} x^{(n+1)\ell} e^{-(n+1)x} L_i^{\ell}(x) L_j^{\ell}(x) dx. \quad (27)$$

This is a special case of the integral shown as Eq. (19) in [24]. Consequently, we obtain the following analytic expression for this integral

$$F_{i,j}^{(n,\ell)} = \frac{\sqrt{(i!)(j!)(i+\ell)!(j+\ell)!}}{\sigma^{\sigma\ell+1}} \times$$

$$\sum_{k=0}^{\min(i,j)} \frac{\sigma^{-2k}}{[(\ell+k)!]^2} \frac{\Gamma(k+\sigma\ell+1)}{k!(i-k)!(j-k)!} {}_2F_1\left(\begin{matrix}k-i, k+\sigma\ell+1\\k+\ell+1\end{matrix}\bigg|\sigma^{-1}\right) {}_2F_1\left(\begin{matrix}k-j, k+\sigma\ell+1\\k+\ell+1\end{matrix}\bigg|\sigma^{-1}\right) \quad (28)$$

where $\sigma := n+1$. All summations in (28), including the hypergeometric series, are finite. Moreover, since for any $n$ and $\ell$, $f^{(n,\ell)}(x)$ is a well-behaved function that vanishes at the origin (except for $\ell = 0$ in which case its value there is one) and decays very rapidly for large $x$, then the integral $F_{i,j}^{(n,\ell)}$ becomes infinitesimal if $i$ or $j$ becomes greater than a large enough integer $N$. On the other hand, a highly accurate evaluation of the integral (26a) could also be achieved using Gauss quadrature that was outlined above where we obtain the following approximation

$$\frac{1}{\ell!} \int_0^{\infty} V\left(\sqrt{x}/\lambda\right) \left[x^{\ell} e^{-x} \tilde{L}_i^{\ell}(x) \tilde{L}_j^{\ell}(x)\right] dx \cong \sum_{l=0}^{N-1} \Lambda_{i,l} W_{l,l} \Lambda_{j,l} = \left(\Lambda W \Lambda^T\right)_{i,j}, \quad (29)$$

where $W_{i,j} = V\left(\sqrt{\xi_i}/\lambda\right) \delta_{i,j}$. Now, the potential function $V(r)$ is non-singular and short-range [i.e., $V(r) \to 0$ for $r > R$ with $R$ being a finite range]. Thus, the quadrature should produce a highly accurate evaluation of this integral even for small enough integer $N$ (i.e., a small $N \times N$ size of the quadrature matrix $J$). Consequently, the matrix elements $(\Lambda W \Lambda^T)_{i,j}$ can be taken zero if either $i$ or $j$ is greater than or equal to $N$ similar to $F_{i,j}^{(n,\ell)}$. The technical notes in Appendix A address the numerical accuracy of evaluating the integrals (26) for large indices $i$ and/or $j$ and how to handle it. Finally, Eq. (25) becomes

$$E A_i^{(m)} = a_i A_i^{(m)} + b_{i-1} A_{i-1}^{(m)} + b_i A_{i+1}^{(m)} + \sum_{j=0}^{N-1}\left(\Lambda W \Lambda^T\right)_{i,j} A_j^{(m)} + g\left(2\lambda^2/\ell!\right)^n \times$$

$$\sum_{j=0}^{N-1} F_{i,j}^{(n,\ell)} \sum_{k_1,k_2,...,k_{2n+1}=0}^{N-1} C_{k_1,k_2,...,k_{2n+1}}^{j} \left(A_{k_1}^{(m-1)} A_{k_2}^{(m-1)} ... A_{k_n}^{(m-1)}\right)\left(\overline{A_{k_{n+1}}^{(m-1)} A_{k_{n+2}}^{(m-1)} ... A_{k_{2n}}^{(m-1)}}\right) A_{k_{2n+1}}^{(m)} \quad (30)$$

Now, we define the elements of the finite $N$-dimensional energy-dependent real symmetric matrix $B^{(m-1)}(E)$ as

$$B_{j,i}^{(m-1)}(E) := \sum_{k_1,k_2,...,k_{2n}=0}^{N-1} C_{k_1,k_2,...,k_{2n},i}^{j} \left(A_{k_1}^{(m-1)} A_{k_2}^{(m-1)} ... A_{k_n}^{(m-1)}\right)\left(\overline{A_{k_{n+1}}^{(m-1)} A_{k_{n+2}}^{(m-1)} ... A_{k_{2n}}^{(m-1)}}\right), \quad (31)$$

for $i, j = 0, 1, ..., N-1$. Then Eq. (30) becomes

$$E A_i^{(m)}(E) = a_i A_i^{(m)}(E) + b_{i-1} A_{i-1}^{(m)}(E) + b_i A_{i+1}^{(m)}(E)$$
$$+ \sum_{j=0}^{N-1}\left(\Lambda W \Lambda^T\right)_{i,j} A_j^{(m)}(E) + g \sum_{j=0}^{N-1} R_{i,j}^{(m-1)}(E) A_j^{(m)}(E) \quad (32)$$

where the energy-dependent matrix elements

$$R_{i,k}^{(m-1)}(E) := \left(2\lambda^2/\ell!\right)^n \sum_{j=0}^{N-1} F_{i,j}^{(n,\ell)} B_{j,k}^{(m-1)}(E). \quad (33)$$

An alternative, but equivalent and numerically more suitable, expression for $R_{i,k}^{(m-1)}(E)$ is given in Appendix A by Eq. (A1).

The only $m$ level unknowns in Eq. (32) are the expansion coefficients $\left\{A_k^{(m)}(E)\right\}_{k=0}^{\infty}$. This algebraic equation (for all $m$) is equivalent to the nonlinear radial Schrödinger equation (10). In matrix notation, it reads

$$E \begin{pmatrix} A_0^{(m)} \\ A_1^{(m)} \\ \times \\ \times \\ \times \\ A_{N-2}^{(m)} \\ A_{N-1}^{(m)} \\ A_N^{(m)} \\ A_{N+1}^{(m)} \\ A_{N+2}^{(m)} \\ \times \\ \times \\ \times \end{pmatrix} = \begin{pmatrix} \times & \times & \times & \times & \times & \times & \times & & & & & & \\ \times & \times & \times & \times & \times & \times & \times & & & & & & \\ \times & \times & \times & \times & \times & \times & \times & & & & & & \\ \times & \times & \times & \times & \times & \times & \times & & & & & & \\ \times & \times & \times & \times & \times & \times & \times & & & & & & \\ \times & \times & \times & \times & \times & \times & \times & & & & & & \\ \times & \times & \times & \times & \times & \times & \times & b_{N-1} & & & & & \\ & & & & & & b_{N-1} & a_N & b_N & & & & \\ & & & & & & & b_N & a_{N+1} & b_{N+1} & & & \\ & & & & & & & & b_{N+1} & a_{N+2} & \times & & \\ & & & & & & & & & \times & \times & \times & \\ & & & & & & & & & & \times & \times & \times \\ & & & & & & & & & & & \times & \times \end{pmatrix} \begin{pmatrix} A_0^{(m)} \\ A_1^{(m)} \\ \times \\ \times \\ \times \\ A_{N-2}^{(m)} \\ A_{N-1}^{(m)} \\ A_N^{(m)} \\ A_{N+1}^{(m)} \\ A_{N+2}^{(m)} \\ \times \\ \times \\ \times \end{pmatrix} \quad (34)$$

The infinite tridiagonal symmetric tail is just the matrix $K$ whose elements are $K_{i,j} = a_i \delta_{i,j} + b_{i-1}\delta_{i-1,j} + b_i \delta_{i+1,j}$ whereas the square block on the top left corner is the $N \times N$ energy-dependent matrix $\left[ K + \Lambda W \Lambda^T + g R^{(m-1)}(E) \right]$. The form of Eq. (34) is identical to that in the linear J-matrix method of scattering [19]. In the following section, we use the tools of the J-matrix method in our perturbative solution but with a modification due to the presence of the nonlinear energy-dependent matrix $R^{(m-1)}(E)$.

## 4. The J-matrix solution

Rows $k = N, N+1, N+2,...$ of Eq. (34) result in the following three-term recursion relation

$$E A_k^{(m)}(E) = a_k A_k^{(m)}(E) + b_{k-1} A_{k-1}^{(m)}(E) + b_k A_{k+1}^{(m)}(E). \tag{35}$$

This recursion relation has two independent solutions, which we call $s_k(E)$ and $c_k(E)$. The general solution, $A_k^{(m)}(E)$, is a linear combination of these with two arbitrary energy-dependent factors that do not depend on the index $k$. These two factors are determined by imposing the boundary conditions. With $\{a_k, b_k\}$ given by Eq. (15), we derive $s_k(E)$ and $c_k(E)$ in Appendix B utilizing a scheme used in the solution of the reference problem in the linear J-matrix method of scattering. These are obtained as

$$s_k(E) = (-1)^k \sqrt{\frac{2(k!)}{\lambda(k+\ell)!}} \mu^{\ell+\frac{1}{2}} e^{-\mu^2/2} L_k^\ell(\mu^2), \tag{36a}$$

$$c_k(E) + i s_k(E) =$$
$$\frac{(-1)^k}{\pi} \sqrt{\frac{2(\ell!)}{\lambda}} \mu^{\ell+\frac{1}{2}} e^{-\mu^2/2} \left[ (-1)^{\ell+1} \Gamma(-\ell, -\mu^2) \tilde{L}_k^\ell(\mu^2) + \sqrt{\frac{1}{\ell+1}} \mu^{-2\ell} e^{\mu^2} \tilde{\mathcal{L}}_{k-1}^\ell(\mu^2; 1) \right] \tag{36b}$$

where $\mu^2 = 2E/\lambda^2$ and $\Gamma(x,y)$ is the upper incomplete gamma function. $\tilde{\mathcal{L}}_k^\ell(z;j)$ is the normalized version of the "associated Laguerre polynomial" of degree $k$ in $z$ and order $j$, which is defined in Appendix B. Imposing the boundary conditions gives the solution of Eq. (35) as follows

$$A_k^{(m)}(E) = \left[ c_k(E) - i s_k(E) \right] - e^{2i\delta_m(E)} \left[ c_k(E) + i s_k(E) \right], \tag{37}$$

for $k = N-1, N, N+1,...$. The phase shift angle $\delta_m(E)$ and the $N-1$ coefficients $\{A_k^{(m)}(E)\}_{k=0}^{N-2}$ are to be determined from the solution of the remaining top $N$ equations in (34) that read

$$\begin{pmatrix} \times & \times & \times & & \times & & \times & \times & \times \\ \times & \times & \times & & \times & & \times & \times & \times \\ \times & \times & \times & & \times & & \times & \times & \times \\ \times & \times & \times & H+gR^{(m-1)}(E)-EI & \times & & \times & \times & \times \\ \times & \times & \times & & \times & & \times & \times & \times \\ \times & \times & \times & & \times & & \times & \times & \times \\ \times & \times & \times & & \times & & \times & \times & \times \end{pmatrix} \begin{pmatrix} A_0^{(m)}(E) \\ A_1^{(m)}(E) \\ \times \\ \times \\ \times \\ A_{N-2}^{(m)}(E) \\ A_{N-1}^{(m)}(E) \end{pmatrix} = \begin{pmatrix} 0 \\ 0 \\ \times \\ \times \\ \times \\ 0 \\ -b_{N-1}A_N^{(m)}(E) \end{pmatrix} \quad (38)$$

where $H$ is the linear component of the $N \times N$ Hamiltonian matrix that reads $H = K + \Lambda W \Lambda^T$ and $I$ is the $N \times N$ unit matrix. We multiply both sides of Eq. (38) by the inverse of this matrix, which is the finite $N \times N$ Green's matrix $G^{(m)}(E) = \left[ H + gR^{(m-1)}(E) - EI \right]^{-1}$, giving

$$\begin{pmatrix} A_0^{(m)}(E) \\ A_1^{(m)}(E) \\ \times \\ \times \\ \times \\ A_{N-2}^{(m)}(E) \\ A_{N-1}^{(m)}(E) \end{pmatrix} = \begin{pmatrix} \times & \times & \times & & \times & & \times & \times & \times \\ \times & \times & \times & & \times & & \times & \times & \times \\ \times & \times & \times & & \times & & \times & \times & \times \\ \times & \times & \times & G^{(m)}(E) & \times & & \times & \times & \times \\ \times & \times & \times & & \times & & \times & \times & \times \\ \times & \times & \times & & \times & & \times & \times & \times \\ \times & \times & \times & & \times & & \times & \times & \times \end{pmatrix} \begin{pmatrix} 0 \\ 0 \\ \times \\ \times \\ \times \\ 0 \\ -b_{N-1}A_N^{(m)}(E) \end{pmatrix} \quad (39)$$

Row $N-1$ of this matrix equation gives a special relation that determines $\delta_m(E)$ by using $A_N^{(m)}(E)$ and $A_{N-1}^{(m)}(E)$ from Eq. (37):

$$e^{2i\delta_m(E)} = \mathcal{T}_{N-1}(E) \frac{1 + b_{N-1} G_{N-1,N-1}^{(m)}(E) \mathcal{R}_N^-(E)}{1 + b_{N-1} G_{N-1,N-1}^{(m)}(E) \mathcal{R}_N^+(E)}, \quad (40)$$

where $\mathcal{T}_k(E) := \frac{c_k(E) - is_k(E)}{c_k(E) + is_k(E)}$, $\mathcal{R}_k^\pm(E) := \frac{c_k(E) \pm is_k(E)}{c_{k-1}(E) \pm is_{k-1}(E)}$, and $G^{(0)}(E) = (H - EI)^{-1}$.

Finally, the remaining coefficients $\{ A_k^{(m)}(E) \}_{k=0}^{N-2}$ are obtained from row 0 to row $N-2$ of Eq. (39) as follows

$$A_k^{(m)}(E) = -b_{N-1} G_{k,N-1}^{(m)}(E) A_N^{(m)}(E), \quad (41)$$

for $k = 0, 1, ..., N-2$. Now the presence of the energy dependent matrix $R^{(m-1)}(E)$ in Eq. (38) makes the evaluation of the finite energy Green's matrix $G^{(m)}(E)$ more elaborate and highly nontrivial as compared to the linear case with $G^{(0)}(E)$. A simple, but tedious procedure, would be to solve Eq. (38) repeatedly at each energy point in a chosen energy range. The challenge is to find a technique that solves it once and for all energies. In Appendix C, we attempt at finding such a technique.

Now with all expansion coefficients $\{ A_k^{(m)}(E) \}_{k=0}^\infty$ being determined via (37) and (41), $\psi_m(r)$ and the scattering matrix $e^{2i\delta_m(E)}$ are well defined. We iterate the procedure to obtain $e^{2i\delta_{m+1}(E)}$, $e^{2i\delta_{m+2}(E)}$, ...etc. until convergence is reached with the desired accuracy. Therefore, we can summarize the steps of the perturbative procedure outlined above to solve the nonlinear Schrödinger equation (10) using the tools of the J-matrix method as follows:

1. Select the physical parameters $\{n, g, \ell\}$ and the non-singular short-range linear potential function $V(r)$.

2. Choose a large enough integer $N$ and an associated value for the non-physical scale parameter $\lambda$. Calculate the eigenvalues $\{\xi_j\}$ and normalized eigenvectors $\{\Lambda_{i,j}\}$ of the $N \times N$ matrix $J$. Observe the diagonal element of the matrices $F^{(n,\ell)}$ and $\Lambda W \Lambda^T$ (within the plateau of stability of $\lambda$) as you reduce $N$. Stop at values of $N$ and $\lambda$ that meet your choice of accuracy.

3. Calculate the linearization coefficients $C^i_{k_1,k_2,...,k_{2n},j}$ using either Eq. (22) or Eq. (23). In the calculation using Eq. (23), take into consideration the technical notes in Appendix A especially the size of the matrix $J$ and the full symmetry. A numerically improved but equivalent alternative is to calculate the tensor $D^{k_1,k_2,...,k_{2n}}_{i,j}$ instead of $C^i_{k_1,k_2,...,k_{2n},j}$ as shown by Eq. (A2) in Appendix A.

4. Calculate the $N \times N$ finite Green's matrix $G^{(0)}(E) = (K + \Lambda W \Lambda^T - EI)^{-1}$ in the sub-basis $\{\phi_k\}_{k=0}^{N-1}$ as shown in Appendix C with $g = 0$.

5. Calculate the expansion coefficients $\{A_k^{(0)}(E)\}_{k=0}^{N}$. These are given as $A_k^{(0)}(E) = [c_k(E) - is_k(E)] - e^{2i\delta_0(E)}[c_k(E) + is_k(E)]$ for $k = N - 1$ and $k = N$ with

$$e^{2i\delta_0(E)} = \mathcal{T}_{N-1}(E) \frac{1 + b_{N-1} G^{(0)}_{N-1,N-1}(E) \mathcal{R}_N^-(E)}{1 + b_{N-1} G^{(0)}_{N-1,N-1}(E) \mathcal{R}_N^+(E)},$$

and $A_k^{(0)}(E) = -b_{N-1} G^{(0)}_{k,N-1}(E) A_N^{(0)}(E)$ for $k = 0, 1, ..., N - 2$.

6. Set $m = 1$.

7. Using $\{A_k^{(m-1)}(E)\}_{k=0}^{N-1}$, calculate the energy-dependent matrix $R^{(m-1)}(E)$ defined by Eq. (33) or preferably by Eq. (A1).

8. Calculate the finite Green's matrix $G_{i,j}^{(m)}(E)$ as shown in Appendix C, then calculate the scattering matrix $e^{2i\delta_m(E)}$ as shown in Eq. (40).

9. If the deviation in the scattering matrix $e^{2i\delta_m(E)}$ from $e^{2i\delta_{m-1}(E)}$ is small enough within your desired accuracy, then exit the procedure otherwise continue.

10. Use equations (37) and (41) to calculate $\{A_k^{(m)}(E)\}_{k=0}^{N}$.

11. Increment $m$ by one, replacing $m$ by $m + 1$ and go to step 7.

In Section 5, we apply this procedure to obtain the perturbative solution for the scattering matrix $e^{2i\delta_m(E)}$ of the cubic ($n = 1$) and quintic ($n = 2$) nonlinearities for a given set of physical parameters. To test the convergence, accuracy and validity of the procedure, we compare the results of the procedure as $g \to 0$ with well-established results of the linear J-matrix calculation for $g = 0$.

In Appendix D, we present an alternative representation for the solution space of the problem but in a basis that differs from the oscillator basis (14). It is referred to as the "Laguerre basis" and has the following elements

$$\phi_k(r) = \sqrt{\frac{\lambda(k!)}{(k+2\ell)!}} (\lambda r)^{\ell+\frac{1}{2}} e^{-\lambda r/2} L_k^{2\ell}(\lambda r) = \sqrt{\frac{\lambda}{(2\ell)!}} (\lambda r)^{\ell+\frac{1}{2}} e^{-\lambda r/2} \tilde{L}_k^{2\ell}(\lambda r). \qquad (42)$$

## 5. The cubic (*n* = 1) and quintic (*n* = 2) nonlinearities

It is well-known that the canonical/regular nonlinearity in 3D is the cubic ($n=1$) nonlinearity, which corresponds to the $\varphi^4$ Lagrangian theory and where the coupling parameter *g* becomes dimensionless. Similar arguments show that the canonical nonlinearity in 2D is quintic ($n=2$), which corresponds to the $\varphi^6$ Lagrangian theory and where *g* becomes dimensionless. Nonetheless, we present in this section results for both nonlinearities in 2D.

We start by validating the 11-step procedure. We do that by comparing the results of the procedure as $g \to 0$ with linear results for $g = 0$. It is not difficult to use the complex scaling method and find the resonance energies associated with the well-known test linear potential $V(r) = 7.5r^2 e^{-r}$ in two dimensions [25]. If we call the energies of such resonances that are associated with the angular momentum $\ell$ as $E_\ell$, then we obtain

$$E_0 = 2.5171 - i0.00024, \qquad E_1 = 4.11 - i0.11. \qquad (43)$$

Using the well-established linear J-matrix method, we can obtain the scattering matrix as a function of energy that clearly shows the location of these resonances [19]. Figure 1 and 2 are plots of $\left|1 - e^{2i\delta_\ell(E)}\right|$ corresponding to $\ell = |\mathrm{m}| = 0$ and $\ell = |\mathrm{m}| = 1$, respectively. The resonance activities associated with the energies (43) are clearly evident in these figures. Now, we calculate the scattering matrix in a tabular form using the 11-step procedure for the physical parameters: $n = 1$, $g = 0.001$, $\ell = 0,1$ and $V(r) = 7.5r^2 e^{-r}$. Table 1 corresponds to $\ell = 0$ and for energy points in the neighborhood of $E = 2.50$ with small increments due to the sharpness of the resonance as indicated by Fig. 1. On the other hand, Table 2 corresponds to $\ell = 1$ and for energy points in the neighborhood of $E = 4.0$ but with larger increments due to the fact that the resonance there is not sharp enough as evident by the plot in Fig. 2. The tables confirm the above findings and is considered a verification of the convergence, accuracy and validity of the 11-step procedure given in Section 4.

Now, we give two examples for calculating the nonlinear scattering matrix $e^{2i\delta(E)}$ (one for $n = 1$ and another for $n = 2$) with a small nonlinear coupling parameter to justify the perturbative nature of calculation in the 11-step procedure. We chose the following non-analytic piece-wise continuous linear potential that traces the analytic function $f(r) = 5r^2 e^{-r}$

$$V(r) = \begin{cases} 2r & 0 \leq r < 1.2 \\ 2.4 & 1.2 \leq r < 3 \\ 4.2 - 0.6r & 3 \leq r < 7 \\ 0 & \text{otherwise} \end{cases} \qquad (44)$$

Table 3 shows the result of calculating $\left|1 - e^{2i\delta_m(E)}\right|$ with the physical parameters: $n=1$, $g=0.02$, and $\ell=1$. We took the computational parameters $\lambda=1$ and $N=20$. Table 4 is a reproduction of Table 3 with the same physical and computational parameters except that the nonlinearity is quintic ($n=2$).

Table 3 indicates convergence of the calculation (for a 6-decimal places accuracy) for all energies up to $m=10$. On the other hand, for the quintic nonlinearity and aside from the two energies at $E=3.0$ and $E=4.0$, Table 4 shows that convergence (for a 6-decimal places accuracy) was reached at $m=9$. Nevertheless, for $E=4.0$ we had to go up to $m=17$ for convergence with the said accuracy to get $\left|1-e^{2i\delta_m(4.0)}\right|=1.945614$. However, a curious but very interesting phenomenon occurs at $E=3.0$ where the result oscillates between two stable values (convergence is reached for large values of $m$). We were able to reach convergence of these two values when $m$ got to 20 but only for a 3-decimal-place accuracy where we obtained $\left|1-e^{2i\delta_m(3.0)}\right|=1.730$ and $\left|1-e^{2i\delta_m(3.0)}\right|=0.075$. This phenomenon is termed "bifurcation" and it is a signature and manifestation of strong nonlinearity. Typically, such bifurcation has a certain number of stable points. In our case, the number of stable points is two.

## 6. Conclusion

The work presented here is our second attempt at an extension of the J-matrix method of scattering to the solution of nonlinear problems. In the first attempt [16], we considered a toy model, excluded an additional linear potential, and proposed an ansatz for the solution. Here, we avoided all those limitations but presented a perturbative formulation where the nonlinear coupling parameter is small. Moreover, due to separability requirement of the original NLSE, we were limited to two-dimensional configuration space. Nonetheless, our solution may prove useful in applications to 2D materials such as graphene, fullerenes, and thin films.

As in the original linear J-matrix method, orthogonal polynomials and Gauss quadrature play important roles in the nonlinear version of the theory. More significantly, is the linearization of products of orthogonal polynomials.

Analysis of the existence and stability of the solution goes beyond the scope of the current work and thus were not performed. Nonetheless, we observed a curious manifestation of the nonlinearity in our results that shows up as bifurcation. The calculation of the nonlinear scattering matrix at certain energy(es) converges to two distinct values oscillating between them.

# Appendix A: Technical notes

In almost all computational software packages (e.g., Mathematica, Python, Mathcad, etc.) adding a number to a matrix result in the addition of that number to all elements of the matrix. However, in the calculation of $C^{j}_{k_1,k_2,...,k_{2n+1}}$ as defined by Eq. (23), the expression $\tilde{L}^{\ell}_k(J) = \alpha_0 + \alpha_1 J + \alpha_2 J^2 + ... + \alpha_k J^k$ should be evaluated as $\tilde{L}^{\ell}_k(J) = \alpha_0 I + \alpha_1 J + \alpha_2 J^2 + ... + \alpha_k J^k$, where $I$ is the unit matrix. Moreover, complete symmetry of $C^{j}_{k_1,k_2,...,k_{2n+1}}$ might not be guaranteed if its definition by Eq. (23) is adopted. Thus, one has to employ numerical routines to insure total symmetry.

The number of elements of the $(2n+2)$ rank tensor $C^{k_{2n+2}}_{k_1,k_2,...,k_{2n+1}}$ is $N^{2n+2}$, where $N$ is the number of all possible values of the index $k_i$. However, due to the symmetry of this tensor under the exchange of any two indices, $k_i$, then the independent elements of this tensor are such that $k_1 \leq k_2 \leq ... \leq k_{2n+1} \leq k_{2n+2}$ or $k_1 \geq k_2 \geq ... \geq k_{2n+1} \geq k_{2n+2}$. This makes the number of independent elements of $C^{k_{2n+2}}_{k_1,k_2,...,k_{2n+1}}$ equals to $(N)_{2n+2}/(2n+2)!$. Further reduction in the calculation cost of this tensor comes about because $C^{k_{2n+2}}_{k_1,k_2,...,k_{2n+1}} = 0$ if the value of any of its indices is greater than the sum of the values of all other indices. Similar arguments hold true for the $(2n+2)$ rank tensor $D^{k_1,k_2,...,k_{2n}}_{i,j}$ defined below by Eq. (A2).

We should note that the linearization of the product of orthogonal polynomials developed in Ref [21], such as the one used to obtain $C^{j}_{k_1,k_2,...,k_{2n+1}}$ in Eq. (23), assumes that the tridiagonal symmetric matrix $J$ is infinite in size. However, for numerical calculations, one is obliged to use a truncated version of the matrix $J$. Now, since the maximum degree of $J$ in the product $\tilde{L}^{\ell}_{k_1}(J)\tilde{L}^{\ell}_{k_2}(J) ... \tilde{L}^{\ell}_{k_{2n}}(J)$ is $k_1 + k_2 + ... + k_{2n}$ then $C^{j}_{k_1,k_2,...,k_{2n+1}}$ contains the element $(k_{2n+1}, j)$ of the finite-size matrix $J^{k_1+k_2+...+k_{2n}}$. Therefore, to avoid numerical errors in the calculation of $C^{j}_{k_1,k_2,...,k_{2n+1}}$ due to this matrix truncation, the size of the truncated matrix $J$ must be larger than $\frac{1}{2}(k_1 + k_2 + ... + k_{2n} + k_{2n+1} + j)$. Assuming that the maximum value of any of the indices of $C^{j}_{k_1,k_2,...,k_{2n+1}}$ is $N-1$, then the size of the matrix $J$ must be larger than or equal to $[(n+1)N - n]$. This numerical error is due to edge-effect caused by missing pieces in matrix multiplication of truncated tridiagonal matrices that are otherwise infinite in size. On the other hand, since the integrand $\tilde{L}^{\ell}_j(x)\tilde{L}^{\ell}_{k_1}(x)\tilde{L}^{\ell}_{k_2}(x) ... \tilde{L}^{\ell}_{k_{2n+1}}(x)$ in the Gauss quadrature integral (19) for $C^{j}_{k_1,k_2,...,k_{2n+1}}$ is a polynomial of degree $j + k_1 + k_2 + ... + k_{2n} + k_{2n+1}$, then according to the Gauss quadrature rule (see page 3 of Ref. [23]), the integral approximation becomes *exact* if the order of the quadrature is greater than $\frac{1}{2}(j + k_1 + k_2 + ... + k_{2n+1})$, which requires that the order of the quadrature (i.e., the size of the matrix $J$) be larger than or equal to $[(n+1)N - n]$.

Numerical evaluation of the integrals (26) and (29) might experience deficiency in the accuracy for large indices $i$ and/or $j$. This is due to the fact that for large degrees, the Laguerre polynomial $\tilde{L}^{\ell}_i(x)$ becomes sinusoidal with very rapid oscillation since $\lim_{i \to \infty} \tilde{L}^{\ell}_i(x) \approx \sin(2\sqrt{ix})$. This can also be seen in the Gauss quadrature approximation of the integral (29) by looking at the diagonal elements of the matrices $\Lambda W \Lambda^T$ where we observe that these values decrease with the matrix size $N$ up to a critical value of $N$ at which they start increasing. We can use our knowledge of the exact formula (28) for the integral (26b) to make an optimum choice of the two numerical parameters (the Gauss quadrature order $N$ and the scale parameter $\lambda$) to obtain a stable, convergent, and accurate result. The objective is to get the most accurate evaluation of the integrals in (26) for the smallest quadrature order by using an optimum value for $\lambda$.

For large $i$ and/or $j$, the numerical evaluation of $F_{i,j}^{(n,\ell)}$ in Eq. (28) may result in floating point error due to the factorials. Alternative, but equivalent, expressions that do not suffer from this numerical shortfall could be derived. For example, numerical evaluation of the integral representation of $F_{i,j}^{(n,\ell)}$ as given by Eq. (27) using Gauss quadrature may circumvent such difficulty. Nonetheless, combing the evaluation of $F_{i,j}^{(n,\ell)}$ and the coefficients $C_{k_1,k_2,\ldots,k_{2n+1}}^j$ into a single Gauss quadrature result to produce the tensor $D_{i,j}^{k_1,k_2,\ldots,k_{2n}}$ shown below in Eq. (A2) could be the best way to eliminate these numerical difficulties.

Finally, during the evaluation of $R_{i,k}^{(m-1)}(E)$ as given by Eq. (33), one may experience two types of difficulties, which are numerical artifacts that arise due to the multiplication of very large numbers by very small numbers: *divergence* and *non-hermiticity*. To eliminate both difficulties simultaneously, we could go back to Eq. (17) and combine the integrals (19) and (26b) into a single Gauss quadrature integral approximation and write

$$R_{i,j}^{(m-1)}(E) = \left(2\lambda^2/\ell!\right)^n \sum_{k_1,k_2,\ldots,k_{2n}=0}^{N-1} D_{i,j}^{k_1,k_2,\ldots,k_{2n}} \left(A_{k_1}^{(m-1)} A_{k_2}^{(m-1)} \ldots A_{k_n}^{(m-1)}\right)\left(\overline{A_{k_{n+1}}^{(m-1)} A_{k_{n+2}}^{(m-1)} \ldots A_{k_{2n}}^{(m-1)}}\right), \quad (A1)$$

where the Gauss quadrature integral approximation (21) gives

$$D_{i,j}^{k_1,k_2,\ldots,k_{2n}} = \int_0^\infty x^{n\ell} e^{-nx} \tilde{L}_{k_1}^\ell(x) \tilde{L}_{k_2}^\ell(x) \ldots \tilde{L}_{k_{2n}}^\ell \left[\left(x^\ell e^{-x}/\ell!\right) \tilde{L}_i^\ell(x) \tilde{L}_j^\ell(x)\right] dx \cong \left[\Lambda Z_{k_1,k_2,\ldots,k_{2n}}^{(n,\ell)} \Lambda^T\right]_{i,j} \quad (A2)$$

with $Z_{k_1,k_2,\ldots,k_{2n}}^{(n,\ell)}$ being an $N \times N$ diagonal matrix whose elements are

$$\begin{aligned}\left[Z_{k_1,k_2,\ldots,k_{2n}}^{(n,\ell)}\right]_{i,j} &= \left[\xi_i^{n\ell} e^{-n\xi_i} \tilde{L}_{k_1}^\ell(\xi_i) \tilde{L}_{k_2}^\ell(\xi_i) \ldots \tilde{L}_{k_{2n}}^\ell(\xi_i)\right] \delta_{i,j} \\ &= \left[\xi_i^{n\ell} e^{-n\xi_i} \Lambda_{k_1,i} \Lambda_{k_2,i} \ldots \Lambda_{k_{2n},i}/(\Lambda_{0,i})^{2n}\right]\delta_{i,j}\end{aligned} \quad (A3)$$

$\{\xi_i\}_{i=0}^{N-1}$ are the eigenvalues of the $N \times N$ matrix $J$ and $\{\Lambda_{j,i}\}_{j=0}^{N-1}$ are the corresponding normalized eigenvectors.

## Appendix B: J-matrix solutions of the recursion relation (35)

The three-term recursion relation (35) for all $k$ corresponds to the linear Schrödinger equation (10) with $g = 0$ and $V(r) = 0$:

$$\mathcal{D}_0 \chi(r) = \left[-\frac{1}{2}\frac{d^2}{dr^2} + \frac{\ell^2 - \frac{1}{4}}{2r^2} - E\right]\chi(r) = 0. \quad (B1)$$

That is, if $h_k(E)$ is a solution of (35), then $\chi(r) = \sum_{k=0}^\infty h_k(E)\phi_k(r)$ is a solution of (B1). It is not difficult to show that Eq. (B1) has two linearly independent solutions, one is regular everywhere while the other blows up at the origin (except for $\ell = 0$). These are written as follows:

$$\chi_{reg}(r) = \sqrt{\kappa r}\,\mathcal{J}_{\ell}(\kappa r)\,, \tag{B2a}$$

$$\chi_{irr}(r) = \sqrt{\kappa r}\,\mathcal{N}_{\ell}(\kappa r)\,, \tag{B2b}$$

where $\kappa = \sqrt{2E}$, $\mathcal{J}_{\ell}(x)$ and $\mathcal{N}_{\ell}(x)$ are the Bessel functions of the first and second kind, respectively (the latter is also known as the Neumann function). The regular solution is energy-normalized, $\langle \chi_{reg} | \chi'_{reg} \rangle = \delta(\kappa - \kappa')$, whereas the irregular solution is not square integrable (with respect to the integration measure $dr$). Near the origin they behave like $\chi_{reg} \to r^{\ell+1/2}$ and $\chi_{irr} \to r^{-\ell+1/2}$. However, asymptotically ($r \to \infty$) both solutions are sinusoidal: $\chi_{reg} \to \sqrt{\frac{2}{\pi}} \cos\left(\kappa r - \ell\frac{\pi}{2} - \frac{\pi}{4}\right)$ and $\chi_{irr} \to \sqrt{\frac{2}{\pi}} \sin\left(\kappa r - \ell\frac{\pi}{2} - \frac{\pi}{4}\right)$. The basis elements (14) are regular everywhere, square-integrable, and vanish at $r = 0$ and $r \to \infty$. Therefore, one of the two solutions of the recursion relation (35), which we refer to as the "sine-like" coefficients $s_k(E)$, gives $\chi(r) = \sum_{k=0}^{\infty} s_k(E)\phi_k(r) := \chi_{\sin}(r)$ that is identical to $\chi_{reg}(r)$ everywhere where $\phi_k(r)$ is given by Eq. (14). The other independent solution, which we call the "cosine-like" coefficients $c_k(E)$, gives $\chi(r) = \sum_{k=0}^{\infty} c_k(E)\phi_k(r) := \chi_{\cos}(r)$ that could be identified with $\chi_{irr}(r)$ only asymptotically. Consequently, $\mathcal{D}_0 \chi_{\sin}(r) = 0$ while $\mathcal{D}_0 \chi_{\cos}(r) \neq 0$. Now, $s_k(E)$ satisfies (35) for $k = 1, 2, ...$ with the initial homogeneous relation $E s_0(E) = a_0 s_0(E) + b_0 s_1(E)$. However, $c_k(E)$ satisfies (35) for $k = 1, 2, ...$ but with the initial non-homogeneous relation $E c_0(E) = a_0 c_0(E) + b_0 c_1(E) - \tau(E)$, for some function $\tau(E)$ to be determined below. The presence of this non-homogeneity is the reason that $\mathcal{D}_0 \chi_{\cos}(r) \neq 0$. Since the solution of the three-term recursion relation (35) is unique up to a factor of an arbitrary function of energy independent of $k$, then we define two independent energy polynomial solutions for (35) and write $s_k(E)$ and $c_k(E)$ as their linear combination with four independent and arbitrary functions of the energy. Let $z = 2E/\lambda^2$ and let us rewrite (35) as

$$z P_k(z) = \eta_k P_k(z) + \sigma_{k-1} P_{k-1}(z) + \sigma_k P_{k+1}(z)\,, \tag{B3}$$

for some polynomials in $z$, $\{P_k(z)\}$, where $\eta_k = 2a_k/\lambda^2$ and $\sigma_k = 2b_k/\lambda^2$. Then $\{p_k(z)\}_{k=0}^{\infty}$ and $\{q_k(z)\}_{k=0}^{\infty}$ are two independent *polynomial* solutions of the recursion relation (B3) for $k = 1, 2, ...$ with the initial values

$$p_0(z) = 1, \qquad p_1(z) = \frac{z - \eta_0}{\sigma_0}, \tag{B4a}$$

$$q_0(E) = 0, \qquad q_1(E) = 1. \tag{B4b}$$

Therefore, $q_k(z)$ becomes a polynomial in $z$ of degree $k-1$, and we can write

$$s_k(E) = \alpha(E)p_k(z) + \alpha'(E)q_k(z), \qquad c_k(E) = \beta(E)p_k(z) + \beta'(E)q_k(E), \tag{B5a}$$

where $\{\alpha, \alpha', \beta, \beta'\}$ are arbitrary and independent functions of $E$. The homogeneous initial relation for $s_k(E)$ gives $\alpha'(E) = 0$ whereas the inhomogeneous initial relation for $c_k(E)$ gives $\beta'(E) = \tau(E)/b_0$. Hence, Eq. (B5a) becomes

$$s_k(E) = \alpha(E) p_k(z), \qquad c_k(E) = \beta(E) p_k(z) + \frac{1}{b_0} \tau(E) q_k(z). \tag{B5b}$$

We start by identifying $p_k(z)$ and $q_k(z)$ then the three energy functions $\alpha(E)$, $\beta(E)$ and $\tau(E)$.

Comparing the recursion relation (35) or (B3) to that of the Laguerre polynomials, one can easily show that $p_k(z) = (-1)^k \tilde{L}_k^\ell(z)$. On the other hand, $q_k(z) = p_{k-1}^{(1)}(z)$, where $p_k^{(j)}(z)$ is called the associated polynomial (a.k.a. the "abbreviated polynomial") of order $j$ (see Section 2.10 of Ref. [22]). It satisfies the same recursion relation (B3) but with $\{\eta_k, \sigma_k\} \mapsto \{\eta_{k+j}, \sigma_{k+j}\}$ and with the initial values $p_0^{(j)}(z) = 1$ and $p_1^{(j)}(z) = (z - \eta_j)/\sigma_j$. Therefore, we can write $q_k(z) = (-1)^{k-1} \tilde{\mathcal{L}}_{k-1}^\ell(z;1)$, where $\tilde{\mathcal{L}}_k^\ell(z;j)$ is the normalized version of the associated Laguerre polynomial $\mathcal{L}_k^\ell(z;j)$ of degree $j$ (see Section 5.6 of Ref. [22]). Now, we turn our attention to the energy functions $\alpha(E)$ and $\beta(E)$. Writing $\chi_{reg}(r) = \sum_{j=0}^\infty s_j(E) \phi_j(r)$ then multiplying both sided by $\phi_i(r)$ and integrating over $r$, we obtain

$$s_i(E) = \sqrt{\frac{2\lambda(i!)}{(i+\ell)!}} \int_0^\infty \sqrt{\kappa r} (\lambda r)^{\ell + \frac{1}{2}} e^{-\lambda^2 r^2/2} L_i^\ell(\lambda^2 r^2) \mathcal{J}_\ell(\kappa r) dr, \tag{B6}$$

where we have used the orthogonality of the Laguerre polynomials on the left-hand side. If we define the dimensionless variable $x = \lambda r$, then the integral (B6) becomes

$$s_i(E) = \frac{1}{\lambda} \sqrt{\frac{2\kappa(i!)}{(i+\ell)!}} \int_0^\infty x^{\ell+1} e^{-x^2/2} L_i^\ell(x^2) \mathcal{J}_\ell(\mu x) dx, \tag{B7}$$

where $\mu = \kappa/\lambda = \sqrt{z}$. This integral is found in Section 7.421 on top of page 812 of Ref. [26] giving

$$s_i(E) = (-1)^i \sqrt{\frac{2(i!)}{\lambda(i+\ell)!}} \mu^{\ell+\frac{1}{2}} e^{-\mu^2/2} L_i^\ell(\mu^2). \tag{B8}$$

Using Eq. (B5b), we can write $s_k(E) = \alpha(E)(-1)^k \tilde{L}_k^\ell(\mu^2) = \alpha(E)(-1)^k \sqrt{\frac{(k!)(\ell!)}{(k+\ell)!}} L_k^\ell(\mu^2)$ giving

$$\alpha(E) = \sqrt{2/\lambda(\ell!)} \mu^{\ell+\frac{1}{2}} e^{-\mu^2/2} = s_0(E). \tag{B9}$$

Using these sine-like coefficients, we plot in Figure 3 $\chi_{sin}(r) = \sum_{k=0}^N s_k(E) \phi_k(r)$ for some large integer $N$ superimposed by $\chi_{reg}(r)$. It is evident that the two plots coincide everywhere as expected. Now we turn to the calculation of $c_k(E)$, which is usually more elaborate.

There are two ways to try to find $c_k(E)$. The first, uses the energy differential equation satisfied by both $s_k(E)$ and $c_k(E)$. The second, is by using the Green's function associated with the operator $\mathcal{D}_0$ and its two independent solutions $\chi_{reg}(r)$ and $\chi_{irr}(r)$. We start with the first method whereby using the expression (B8) for $s_k(E)$ and the differential equation of the

–17–

Laguerre polynomials, we can write the following 2nd order differential equation (with $z = \mu^2$)

$$\left[ z\frac{d^2}{dz^2} + \frac{1}{2}\frac{d}{dz} - \frac{z}{4} - \frac{\ell^2 - \frac{1}{4}}{4z} + \frac{1}{2}(2k+\ell+1) \right] s_k(E) = 0. \tag{B10}$$

which must also be satisfied by $c_k(E)$. Writing the general solution of (B10) as $z^\nu e^{-z/2} F(z)$, we obtain

$$\left[ z\frac{d^2}{dz^2} + \left(2\nu + \tfrac{1}{2} - z\right)\frac{d}{dz} + \frac{(2\nu - \tfrac{1}{2})^2 - \ell^2}{4z} + \frac{1}{2}\left(2k - 2\nu + \ell + \tfrac{1}{2}\right) \right] F(z) = 0. \tag{B11}$$

Choosing $2\nu = \tfrac{1}{2} \pm \ell$, this equation becomes

$$\left[ z\frac{d^2}{dz^2} + (1 \pm \ell - z)\frac{d}{dz} + \left(k + \tfrac{1 \mp 1}{2}\ell\right) \right] F_\pm(z) = 0. \tag{B12}$$

Comparing this to the confluent hypergeometric (Kummer) differential equation (see Section 6.3 of Ref. [27]), we find that $F_+(z) = {}_1F_1(-k; 1+\ell; \mu^2)$ and $F_-(z) = U(-k-\ell; 1-\ell; \mu^2)$, where $U(a;b;z) = \frac{\Gamma(1-b)}{\Gamma(1+a-b)}{}_1F_1(a;b;z) + \frac{\Gamma(b-1)}{\Gamma(a)} z^{1-b} {}_1F_1(1+a-b; 2-b; z)$. It is obvious that the solution associated with $F_+(z)$ is $\mu^{\ell+\frac{1}{2}} e^{-\mu^2/2} {}_1F_1(-k; 1+\ell; \mu^2)$, which is proportional to $s_k(E)$ with a proportionality that depends on $k$ and $\ell$. However, the second solution associated with $F_-(z)$ is not acceptable because $U(-k-\ell; 1-\ell; \mu^2)$ satisfies the recursion relation (35) with the homogeneous initial relation, which is not compatible with $c_k(E)$. Now, there are other solutions of (B12) listed in Section 6.3.1 of Ref. [27] that could be studied to see whether any of them satisfy the non-homogeneous recursion relation. However, those solutions are too elaborate involving the log function, infinite terms each with a sum, or the Euler psi-function. Therefore, we opt for the Green's function approach.

Using Eq. (13), we can write $\mathcal{D}_0 \chi_{\cos}(r) = \sum_{k=0}^\infty \phi_k(r)\left[(a_k - E)c_k + b_{k-1}c_{k-1} + b_k c_{k+1}\right]$ and due to the non-homogeneity in the recursion relation for $\{c_k\}$, only the zero term in the sum survives giving

$$\mathcal{D}_0 \chi_{\cos}(r) = \phi_0(r)\left[(a_0 - E)c_0 + b_0 c_1\right] = \tau(E)\phi_0(r). \tag{B13}$$

Now, the Green's function $G(r,r')$ associated with the operator $\mathcal{D}_0$ satisfies $\mathcal{D}_0 G(r,r') = -\delta(r-r')$ and could be written in terms of its two independent solutions as

$$G(r,r') = 2\left[\chi_{reg}(r_<)\chi_{irr}(r_>)\right]/W(\chi_{reg}, \chi_{irr}), \tag{B14}$$

where $r_>$ and $r_<$ stands for the largest and smallest of $r$ and $r'$. $W(\chi_{reg}, \chi_{irr})$ is the Wronskian of the two solutions (B2), which can be evaluated using the Wronskian of the Bessel functions (see, for example, Section 3.1.1 of [27]) giving $W(\chi_{reg}, \chi_{irr}) = 2\kappa/\pi = 2\lambda\mu/\pi$. Therefore, using (B13) and the Green's function, we can write

$$\chi_{\cos}(r) = -\tau(E)\int_0^\infty G(r,r')\phi_0(r')dr'$$

$$= -\frac{2\tau(E)}{W(E)}\left[\chi_{irr}(r)\int_0^r \chi_{reg}(r')\phi_0(r')dr' + \chi_{reg}(r)\int_r^\infty \chi_{irr}(r'))\phi_0(r')dr'\right] \quad (B15)$$

Using $\chi_{reg}(r) = \chi_{\sin}(r)$ and imposing the boundary condition $\lim_{r\to\infty}[\chi_{\cos}(r) = \chi_{irr}(r)]$, we obtain

$$\tau(E) = -W(E)/2s_0(E) = -(\lambda/\pi)\sqrt{\lambda(\ell!)/2}\,\mu^{-\ell+\frac{1}{2}}e^{\mu^2/2}. \quad (B16)$$

To determine the remaining arbitrary function $\beta(E)$, we apply the energy differential equation (B10) on $c_0(E)$, which is just $\beta(E)$:

$$\left[z\frac{d^2}{dz^2} + \frac{1}{2}\frac{d}{dz} - \frac{z}{4} - \frac{\ell^2 - \frac{1}{4}}{4z} + \frac{1}{2}(\ell+1)\right]c_0(E) = 0. \quad (B17)$$

Following the same steps as in the solution of Eq. (B10) above, we write $c_0(E) = z^{\frac{\ell}{2}+\frac{1}{4}}e^{-z/2}F(z)$ giving

$$\left[z\frac{d}{dz} + (1+\ell-z)\right]F'(z) = 0, \quad (B18)$$

where $F'(z) = dF/dz$. The solution of this equation is $F'(z) \propto z^{-\ell-1}e^z$. Thus, $F(z) \propto \int z^{-\ell-1}e^z dz = (-1)^{\ell+1}\Gamma(-\ell,-z)$ where $\Gamma(x,y)$ is the upper incomplete gamma function. Therefore, we can write

$$c_0(E) = \frac{Q_\ell}{\sqrt{\lambda}}(-1)^{\ell+1}\mu^{\ell+\frac{1}{2}}e^{-\mu^2/2}\Gamma(-\ell,-\mu^2), \quad (B19)$$

where we have written the proportionality factor as $Q_\ell/\sqrt{\lambda}$ with $Q_\ell$ depending only on $\ell$ and we inserted $1/\sqrt{\lambda}$ for correct dimensionality. The following relation is well-known (see, for example, Chapter IX of Ref. [27])

$$\Gamma(-\ell,x) = \frac{1}{\ell!}\left[\frac{e^{-x}}{x^\ell}\sum_{m=0}^{\ell-1}(\ell-m-1)!(-x)^m + (-1)^\ell\Gamma(0,x)\right], \quad (B20)$$

and where $\Gamma(0,x) = -\gamma - \ln(x) - \sum_{m=1}^\infty\left[(-x)^m/m(m!)\right]$ with $\gamma$ being the Euler-Mascheroni constant[1]. Consequently, $c_0(E)$ becomes a complex function due to the negative real argument of the log function in $\Gamma(0,-\mu^2)$. Now, the function $c_k(E)$ must be real as is $s_k(E)$. Therefore, we should investigate the real and imaginary parts of the solution (B19). Now, the imaginary part of $\ln(-\mu^2)$ is equal to $\pi$ independent of $\mu$. Therefore, $\text{Im}\,c_0(E) = \frac{Q_\ell}{\sqrt{\lambda}}(-1)^{\ell+1}\mu^{\ell+\frac{1}{2}}e^{-\mu^2/2} \times$ $\left[ -(-1)^\ell \pi/\ell! \right]$, which is identical to $s_0(E)$ if we choose $Q_\ell = \sqrt{2(\ell!)}/\pi$. Hence, we conclude that

$$c_0(E) = \beta(E) = \frac{(-1)^{\ell+1}}{\pi}\sqrt{\frac{2(\ell!)}{\lambda}}\mu^{\ell+\frac{1}{2}}e^{-\mu^2/2}\operatorname{Re}\left[\Gamma(-\ell,-\mu^2)\right]. \tag{B21a}$$

$$s_0(E) = \alpha(E) = \frac{(-1)^{\ell+1}}{\pi}\sqrt{\frac{2(\ell!)}{\lambda}}\mu^{\ell+\frac{1}{2}}e^{-\mu^2/2}\operatorname{Im}\left[\Gamma(-\ell,-\mu^2)\right]. \tag{B21b}$$

---

[1] The numerical value of Euler-Mascheroni constant, to 50 decimal places, is: 0.57721566490153286060651209008240243104215933593992...

To test the validity of these findings, we use $c_0(E)$ from (B21a) and $\tau(E)$ from (B16) into the initial relation of the inhomogeneous recursion to get $c_1(E) = \left[(E-a_0)c_0(E) + \tau(E)\right]/b_0$. Then, we use the three-term recursion relation (35) or its equivalent (B3) to obtain $c_k(E)$ for $k = 2,3,4,..$ starting with the initial values $c_0(E)$ and $c_1(E)$. Using these cosine-like coefficients, we plot in Figure 4 $\chi_{\cos}(r) = \sum_{k=0}^{N} c_k(E)\phi_k(r)$ for some large integer $N$ superimposed by $\chi_{irr}(r)$. It is evident from the figure that the two plots are matched asymptotically as expected. Finally, we can use (B5b) to write

$$c_k(E) = \frac{\beta(E)}{\alpha(E)}s_k(E) + \frac{(-1)^{k-1}}{b_0}\tau(E)\tilde{\mathcal{L}}_{k-1}^\ell(z;1)$$

$$= (-1)^{\ell+1}\frac{\ell!}{\pi}\operatorname{Re}\left[\Gamma(-\ell,-\mu^2)\right]s_k(E) + \frac{(-1)^k}{\pi}\sqrt{\frac{2(\ell!)}{\lambda(\ell+1)}}\mu^{-\ell+\frac{1}{2}}e^{\mu^2/2}\tilde{\mathcal{L}}_{k-1}^\ell(z;1) \tag{B22}$$

Alternatively, we can combine the J-matrix coefficients as follows:

$$c_k(E) + is_k(E) =$$
$$\frac{(-1)^k}{\pi}\sqrt{\frac{2(\ell!)}{\lambda}}\mu^{\ell+\frac{1}{2}}e^{-\mu^2/2}\left[(-1)^{\ell+1}\Gamma(-\ell,-\mu^2)\tilde{L}_k^\ell(\mu^2) + \sqrt{\frac{1}{\ell+1}}\mu^{-2\ell}e^{\mu^2}\tilde{\mathcal{L}}_{k-1}^\ell(\mu^2;1)\right]. \tag{B23}$$

## Appendix C: Evaluation of the Green's matrix $G^{(m)}(z)$

In this Appendix, we present a procedure for evaluating the finite Green's matrix $G^{(m)}(E)$ needed to solve the matrix equation (38). A tedious procedure would be to solve the equation repeatedly at each energy point within a desired energy range where in equations (C9a-C9d) below we make the replacement: $\Omega \mapsto I$, $\tilde{H} \mapsto K + \Lambda W \Lambda^T + g R^{(m-1)}(E)$ and $E^{(m)} \mapsto E$. In that case, the first step in the 11-step procedure of Section 4 should be changed to read "Select the physical parameters $\{n,g,\ell,E\}$ and the…". Then, the whole procedure will be repeated for several values of $E$ within a desired energy range. Nonetheless, here we propose a procedure to solve it once and for all energies.

We start by making a basis transformation of the "inner" function space spanned by $\{\phi_n(x)\}_{n=0}^{N-1}$ as $\phi_n(x) \mapsto \sum_{k=0}^{N-1}\Omega_{k,n}\phi_k(x)$ where $\Omega$ is an $N \times N$ energy-independent real matrix to be determined. The wavefunction expansion shows that this is equivalent to $A_n^{(m)}(E) \mapsto$ $\sum_{n=0}^{N-1} \Omega_{k,n} A_n^{(m)}(E)$, which in matrix notation it reads $|A^{(m)}\rangle \mapsto \Omega |A^{(m)}\rangle$. Therefore, under this transformation the left-hand side of Eq. (38) becomes $\left[ H + g \hat{R}^{(m-1)}(E) - EI \right] \Omega |A^{(m)}\rangle$, where $\hat{R}^{(m-1)}(E)$ is defined below by Eq. (C1). On the other hand, the right-hand side remains the same since $A_N^{(m)}(E)$ is not affected by this transformation. Now, if we multiply the transformed matrix equation (38) from left by $\Omega^{-1}$, we obtain the following matrix transformations: $H \mapsto \tilde{H} = \Omega^{-1} H \Omega$ and $I \mapsto \tilde{I} = \Omega^{-1} I \Omega = I$. However, the nonlinear energy-dependent matrix $R^{(m-1)}(E)$ defined in Appendix A by Eq. (A1) transforms as $R^{(m-1)} \mapsto \tilde{R}^{(m-1)} = \Omega^{-1} \hat{R}^{(m-1)} \Omega$ where:

$$\hat{R}_{i,j}^{(m-1)}(E) = \left( 2\lambda^2 / \ell! \right)^n \sum_{k_i, k_i' = 0}^{N-1} \left\{ D_{i,j}^{k_1, k_2, \ldots, k_{2n}} \Omega_{k_1, k_1'} \Omega_{k_2, k_2'} \ldots \Omega_{k_{2n}, k_{2n}'} \times \right. \\ \left. \left[ A_{k_1'}^{(m-1)}(E) A_{k_2'}^{(m-1)}(E) \ldots A_{k_n'}^{(m-1)}(E) \right] \left[ \overline{A_{k_{n+1}'}^{(m-1)}(E) A_{k_{n+2}'}^{(m-1)}(E) \ldots A_{k_{2n}'}^{(m-1)}(E)} \right] \right\} \quad (C1)$$

At this point, we require that for a given $n$ and $\ell$, the transformation matrix $\Omega$ must satisfy

$$\sum_{k_1, k_2, \ldots, k_{2n} = 0}^{N-1} D_{i,j}^{k_1, k_2, \ldots, k_{2n}} \Omega_{k_1, k_1'} \Omega_{k_2, k_2'} \ldots \Omega_{k_{2n}, k_{2n}'} = T_{i,j} \delta_{k_1', k_2', \ldots, k_{2n}'}, \quad (C2)$$

where $T$ is an $N \times N$ symmetric matrix to be determined and $\delta_{k_1', k_2', \ldots, k_{2n}'}$ is the totally symmetric $2n$-dimensional Kronecker delta such that for any tensor $P$ of rank $2n$, we can write

$$\sum_{k_1, k_2, \ldots, k_{2n} = 0}^{N-1} P_{k_1, k_2, \ldots, k_{2n}} \delta_{k_1, k_2, \ldots, k_{2n}} = \sum_{k=0}^{N-1} P_{k, k, \ldots, k}. \quad (C3)$$

Thus, we obtain $\hat{R}_{i,j}^{(m-1)}(E) = \left( 2\lambda^2 / \ell! \right)^n \omega^{(m-1)}(E) T_{i,j}$ where the energy dependent function $\omega^{(m-1)}(E)$ reads

$$\omega^{(m-1)}(E) = \sum_{k=0}^{N-1} \left| A_k^{(m-1)}(E) \right|^{2n}. \quad (C4)$$

Hence, the nonlinear energy-dependent matrix $R^{(m-1)}(E)$ transforms as follows:

$$R_{i,j}^{(m-1)}(E) \mapsto \tilde{R}_{i,j}^{(m-1)}(E) = \left( 2\lambda^2 / \ell! \right)^n \omega^{(m-1)}(E) \left( \Omega^{-1} T \Omega \right)_{i,j}. \quad (C5)$$

Finally, we impose the following constraint on the matrices $T$ and $\Omega$

$$\Omega^{-1} T \Omega = I, \quad (C6)$$

A simple choice is that $T_{i,j} = \delta_{i,j}$. Putting all of the above together, Eq. (38) becomes

$$\begin{pmatrix} \times & \times & \times & & \times & & \times & \times & \times \\ \times & \times & \times & & \times & & \times & \times & \times \\ \times & \times & \times & & \times & & \times & \times & \times \\ \times & \times & \times & \tilde{H}-E^{(m)}I & \times & & \times & \times & \times \\ \times & \times & \times & & \times & & \times & \times & \times \\ \times & \times & \times & & \times & & \times & \times & \times \\ \times & \times & \times & & \times & & \times & \times & \times \end{pmatrix} \begin{pmatrix} A_0^{(m)}(E) \\ A_1^{(m)}(E) \\ \times \\ \times \\ \times \\ A_{N-2}^{(m)}(E) \\ A_{N-1}^{(m)}(E) \end{pmatrix} = -b_{N-1}A_N^{(m)}(E) \begin{pmatrix} \Omega_{0,N-1}^{-1} \\ \Omega_{1,N-1}^{-1} \\ \times \\ \times \\ \times \\ \Omega_{N-2,N-1}^{-1} \\ \Omega_{N-1,N-1}^{-1} \end{pmatrix} \quad (C7)$$

where $E^{(m)} = E - g\left(2\lambda^2/\ell!\right)^n \omega^{(m-1)}(E)$ with $E^{(0)} = E$. Consequently, the scattering matrix equation (40) gets replaced by:

$$e^{2i\delta_m(E)} = \mathcal{T}_{N-1}(E) \frac{1 + b_{N-1}\tilde{G}_{N-1,N-1}^{(m)}(E)\mathcal{R}_N^-(E)}{1 + b_{N-1}\tilde{G}_{N-1,N-1}^{(m)}(E)\mathcal{R}_N^+(E)}, \quad (C8)$$

where $\tilde{G}_{j,N-1}^{(m)}(E) = \sum_{k=0}^{N-1} G_{j,k}^{(m)}(E)\Omega_{k,N-1}^{-1}$. Moreover, Eq. (41) gets replaced by: $A_k^{(m)}(E) = -b_{N-1}\tilde{G}_{k,N-1}^{(m)}(E)A_N^{(m)}(E)$ for $k = 0,1,...,N-2$.

Now, the evaluation of the finite energy Green's matrix $G^{(m)}(E^{(m)}) := \left(\tilde{H} - E^{(m)}I\right)^{-1}$ in the orthonormal basis $\{\tilde{\phi}_k\}_{k=0}^{N-1}$, where $|\tilde{\phi}\rangle = \Omega^T|\phi\rangle$, could be written in one of two ways as follows [28]:

$$G_{i,j}^{(m)}(z) = \sum_{k=0}^{N-1} \frac{\Gamma_{i,k}\Gamma_{j,k}}{\varepsilon_k - z}, \quad (C9a)$$

$$G_{i,j}^{(m)}(z) = (-1)^{i+j} \frac{\prod_{k=0}^{N-2} \varepsilon_k^{(i,j)}(z)}{\prod_{k=0}^{N-1} \varepsilon_k - z}, \quad (C9b)$$

where $\{\varepsilon_k\}_{k=0}^{N-1}$ are the eigenvalues of the $N \times N$ matrix $\tilde{H}$ and $\{\Gamma_{i,k}\}_{i=0}^{N-1}$ are the corresponding normalized eigenvectors. The quantities $\{\varepsilon_k^{(i,j)}(z)\}_{k=0}^{N-2}$ are the eigenvalues of an $(N-1) \times (N-1)$ abbreviated matrix obtained from $(\tilde{H} - zI)$ by deleting row $i$ and column $j$. An alternative but equivalent formula to (C9b) with a substantial reduction in the computational cost, which is due to the functional dependence on $z$ of $\varepsilon_k^{(i,j)}(z)$, could be written as follows [28]

$$G_{i,j}^{(m)}(z) = (-1)^{i+j} \sum_{k=0}^{N-1} \frac{\prod_{m=0}^{N-2}\varepsilon_{m,k}^{(i,j)}}{(\varepsilon_k - z)\prod_{m\neq k}^{N-1}(\varepsilon_m - \varepsilon_k)}, \quad (C9c)$$

where $\{\varepsilon_{m,k}^{(i,j)} = \varepsilon_m^{(i,j)}(\varepsilon_k)\}_{m=0}^{N-2}$ are the eigenvalues of an $(N-1) \times (N-1)$ abbreviated matrix obtained from $(\tilde{H} - \varepsilon_k I)$ by deleting row $i$ and column $j$. Numerically, formula (C9c) is more preferable over (C9a) since it does not require the sumptuous evaluation of eigenvectors. For the diagonal elements of the Green's matrix, formula (C9c) simplifies to read

$$G_{i,i}^{(m)}(z) = \frac{\prod_{k=0}^{N-2}\varepsilon_k^{(i,i)} - z}{\prod_{k=0}^{N-1}\varepsilon_k - z} = \sum_{k=0}^{N-1}\frac{\Gamma_{i,k}^2}{\varepsilon_k - z}, \tag{C9d}$$

where $\{\varepsilon_k^{(i,i)}\}_{k=0}^{N-2}$ are the eigenvalues of an $(N-1)\times(N-1)$ abbreviated matrix obtained from $\tilde{H}$ by deleting row $i$ and column $i$.

## Appendix D: J-matrix solution of the reference problem in the Laguerre basis

In this Appendix, we present an alternative but equivalent solution of the reference problem [i.e., obtain $s_k(E)$ and $c_k(E)$] in the Laguerre basis whose elements are shown in Eq. (42). Such basis is numerically more suitable for problem with longer range potentials $V(r)$. This is due to the fact that the basis elements involve the weaker exponentially decaying factor $e^{-\lambda r/2}$ as compared to the oscillator basis of Eq. (14) where it is $e^{-\lambda^2 r^2/2}$. Thus, such basis elements will result in better sampling of the interaction at distances farther away from the origin compared to those of the oscillator basis that work better at distances closer to the origin. That is why in nuclear scattering, which is short in range, calculation is mostly carried out using the oscillator basis whereas in atomic scattering, which is long-range, is computed using the Laguerre basis.

In the Laguerre basis (42), the matrix elements of the free wave operator $\mathcal{D}_0 = -\frac{1}{2}\frac{d^2}{dr^2} + \frac{\ell^2 - \frac{1}{4}}{2r^2} - E$ are

$$\langle\phi_i|\mathcal{D}_0|\phi_j\rangle = \frac{\lambda^2}{2}\left(\mu^2 + \tfrac{1}{4}\right)\times \\ \left[-2\left(i+\ell+\tfrac{1}{2}\right)(\cos\theta)\delta_{i,j} + \sqrt{i(i+2\ell)}\,\delta_{i,j+1} + \sqrt{(i+1)(i+2\ell+1)}\,\delta_{i,j-1}\right] \tag{D1}$$

where $\mu = \sqrt{2E}/\lambda$, $\cos\theta = \frac{\mu^2 - \frac{1}{4}}{\mu^2 + \frac{1}{4}}$ with $0 < \theta \le \pi$, and $\sin\theta = \frac{\mu}{\mu^2 + \frac{1}{4}}$. Therefore, the expansion coefficients $s_k(E)$ and $c_k(E)$ must satisfy the following three-term recursion relation

$$2\left(k+\ell+\tfrac{1}{2}\right)(\cos\theta)P_k(E) = \sqrt{k(k+2\ell)}\,P_{k-1}(E) + \sqrt{(k+1)(k+2\ell+1)}\,P_{k+1}(E), \tag{D2}$$

but with two different initial values for $P_0(E)$ and $P_1(E)$. Following a procedure similar to that in Appendix B for the oscillator basis, we obtain

$$s_k(E) = \frac{2^\ell}{\sqrt{\pi\lambda}}\Gamma\left(\ell + \tfrac{1}{2}\right)(\sin\theta)^{\ell+\frac{1}{2}}\sqrt{\frac{k!}{(k+2\ell)!}}\,C_k^{\ell+\frac{1}{2}}(\cos\theta), \tag{D3}$$

$$c_k(E) = \frac{2^{\ell+1}\ell!}{\pi\sqrt{\lambda}}(\sin\theta)^{\ell+\frac{1}{2}}\sqrt{\frac{k!}{(k+2\ell)!}} \times \qquad (D4)$$
$$\left[(\cos\theta)\,_2F_1\left(\tfrac{1}{2},\ell+1;\tfrac{3}{2};\cos^2\theta\right)C_k^{\ell+\frac{1}{2}}(\cos\theta) - (\sin\theta)^{-2\ell}\mathcal{C}_{k-1}^{\ell+\frac{1}{2}}(\cos\theta)\right]$$

where $C_k^\nu(x)$ is the conventional Gegenbauer (ultra-spherical) polynomial whereas $\mathcal{C}_k^\nu(x)$ is an associated Gegenbauer polynomial that satisfies the following recursion relation

$$2(k+\nu+1)x\mathcal{C}_k^\nu(x) = (k+2)\mathcal{C}_{k+1}^\nu(x) + (k+2\nu)\mathcal{C}_{k-1}^\nu(x), \qquad (D5)$$

with $\mathcal{C}_{-1}^\nu(x) := 0$, $\mathcal{C}_0^\nu(x) = 1$, and $\mathcal{C}_1^\nu(x) = (\nu+1)x$. Figures 5 is a reproduction of Figures 3(b) but in the Laguerre basis. Similarly, is Figure 6 with Figure 4(b).

## Figure Captions

**Fig. 1**: Plot of the linear scattering matrix, $\left|1-e^{2i\delta_0(E)}\right|$, as a function of the energy associated with the potential $V(r)=7.5r^2 e^{-r}$ for $\ell=|\mathsf{m}|=0$ using the linear J-matrix method. The plot shows clearly the resonance activity at $E_0=2.517$.

**Fig. 2**: Plot of the linear scattering matrix, $\left|1-e^{2i\delta_1(E)}\right|$, as a function of the energy associated with the potential $V(r)=7.5r^2 e^{-r}$ for $\ell=|\mathsf{m}|=1$ using the linear J-matrix method. The plot shows clearly the resonance activity at $E_1=4.11$.

**Fig. 3**: Plots of $\chi_{\sin}(r)$ (red solid trace) and $\chi_{reg}(r)$ (blue dotted trace) for several values of the energy (in atomic units) and $\ell$: (a) $\ell=0$ and $E=1.5$, (b) $\ell=1$ and $E=1.0$, (c) $\ell=2$ and $E=1.5$, (d) $\ell=3$ and $E=2.5$. In all these plots, we took $\lambda=1.0$ and $N=1000$. The two black dashed horizontal lines are at $\pm\sqrt{2/\pi}$. The radial distance is measured in units of $\lambda^{-1}$.

**Fig. 4**: Plots of $\chi_{\cos}(r)$ (red solid trace) and $\chi_{irr}(r)$ (blue dashed trace) for several values of the energy (in atomic units) and $\ell$: (a) $\ell=0$ and $E=1.5$, (b) $\ell=1$ and $E=1.0$, (c) $\ell=2$ and $E=1.5$, (d) $\ell=3$ and $E=2.5$. In all these plots, we took $\lambda=1.0$ and $N=1000$. The two black dashed horizontal lines are at $\pm\sqrt{2/\pi}$. The radial distance is measured in units of $\lambda^{-1}$.

**Fig. 5**: Reproduction of Figure 3(b) but in the Laguerre basis with the same parameters.

**Fig. 6**: Reproduction of Figure 4(b) but in the Laguerre basis with the same parameters.

## Table Captions

**Table 1**: Evaluation of $\left|1-e^{2i\delta_m(E)}\right|$ using the 11-step procedure of Section 4 with the physical parameters $\{n=1, g=0.001, \ell=0\}$ and $V(r)=7.5r^2 e^{-r}$. The Table indicates convergence of the calculation (for a 6-decimal places accuracy) up to $m=7$ and clearly shows a resonance activity at $E_0=2.50$.

**Table 2**: Evaluation of $\left|1-e^{2i\delta_m(E)}\right|$ using the 11-step procedure of Section 4 with the physical parameters $\{n=1, g=0.001, \ell=1\}$ and $V(r)=7.5r^2 e^{-r}$. The Table indicates convergence of the calculation (for a 6-decimal places accuracy) up to $m=7$ and clearly shows a resonance activity at $E_1=4.1$.

**Table 3**: Evaluation of $\left|1-e^{2i\delta_m(E)}\right|$ using the 11-step perturbative procedure of Section 4 with the physical parameters $\{n=1, g=0.02, \ell=1\}$ and $V(r)$ as given by Eq. (44). We took the computational parameters $\lambda=1$, $N=20$, and a Gauss quadrature of order 100. The Table indicates convergence of the calculation (for a 6-decimal places accuracy) up to $m=10$.

**Table 4**: Reproduction of Table 3 but for the quintic nonlinearity ($n=2$). Due to numerical limitations, we took a Gauss quadrature of order 30 but kept $N=20$. For convergence and stability of the calculation in this case, please refer to the discussion in the last paragraph of Section 5.

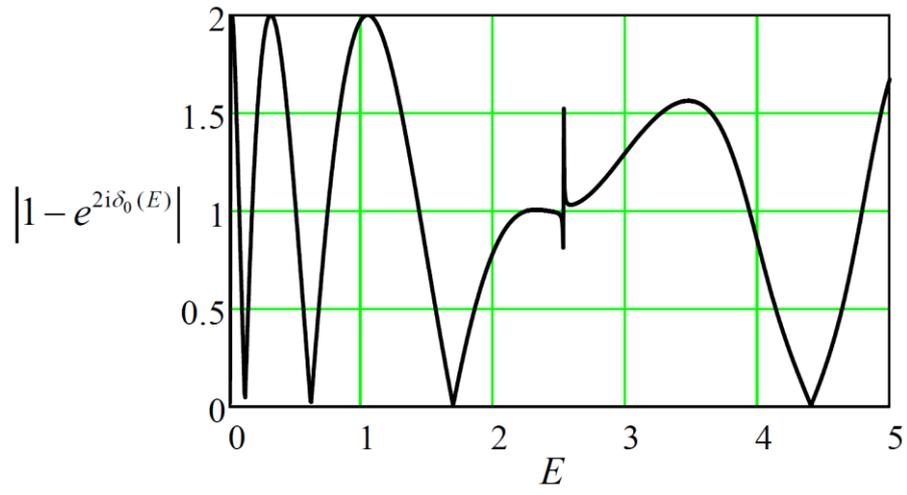

**Fig. 1**

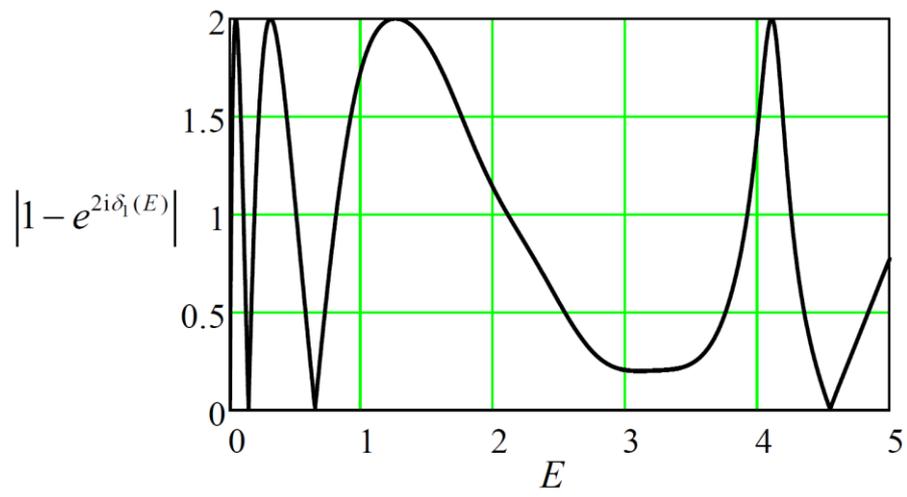

**Fig. 2**

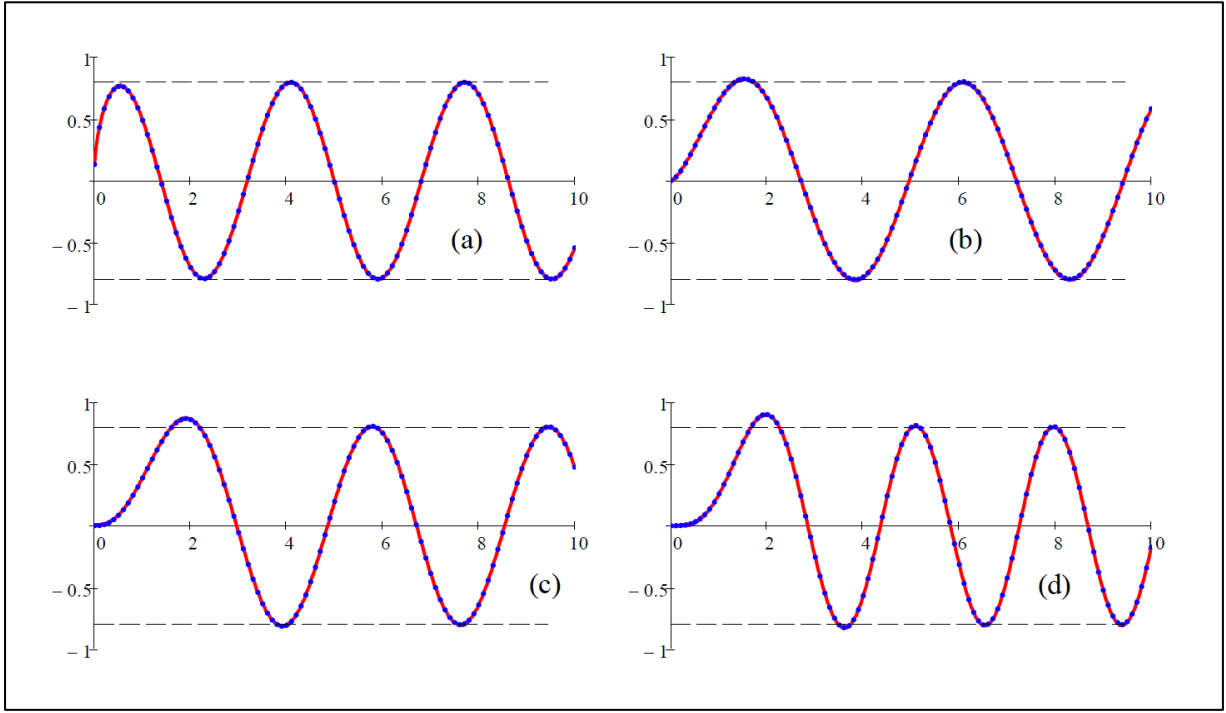

**Fig. 3**

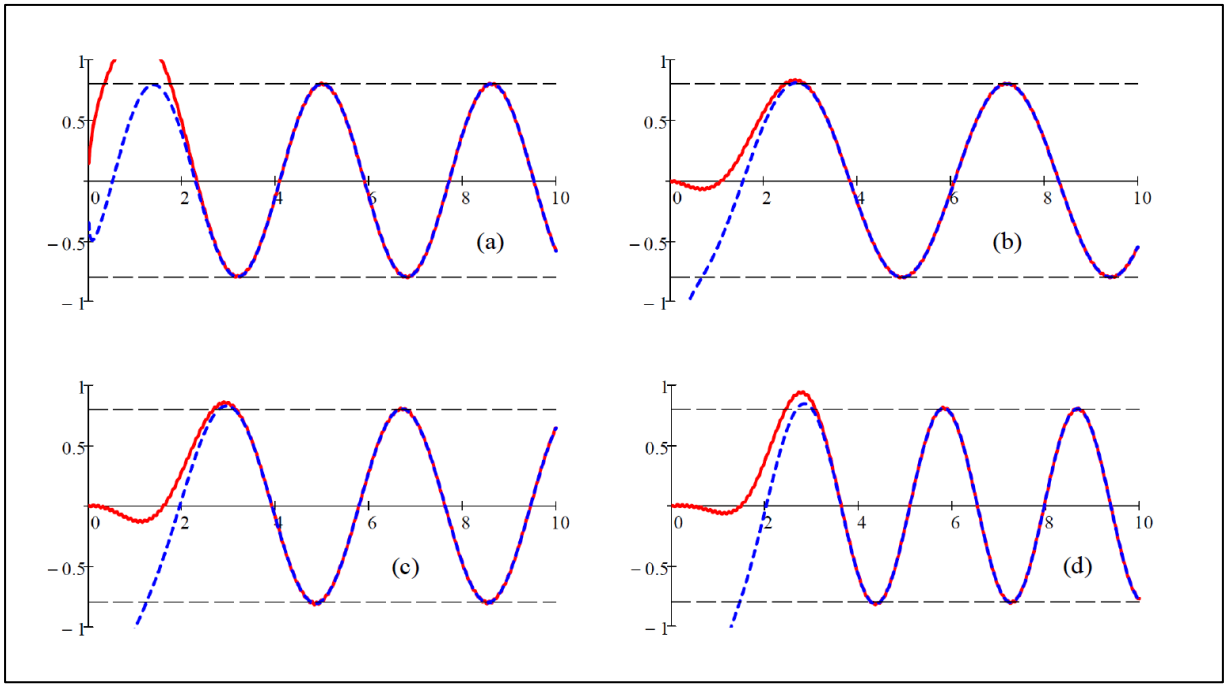

**Fig. 4**

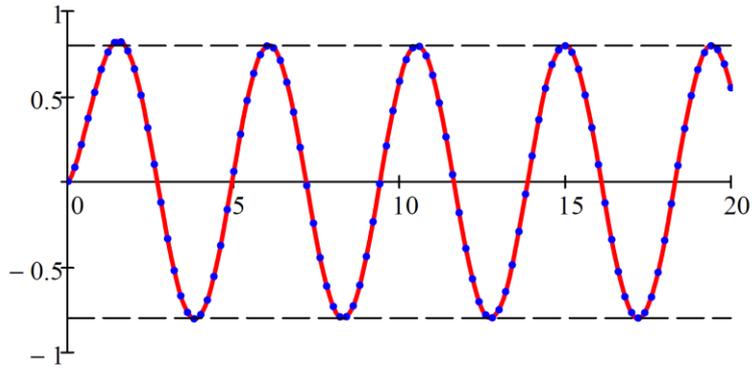

**Fig. 5**

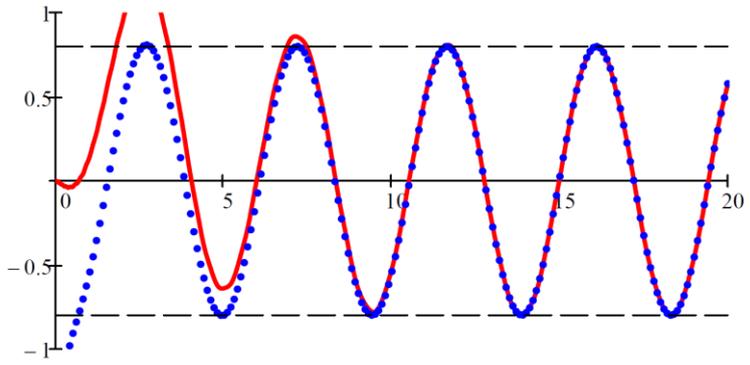

**Fig. 6**

**Table 1**

|       | $E = 2.40$ | $E = 2.45$ | $E = 2.50$ | $E = 2.55$ | $E = 2.60$ |
|-------|-----------|-----------|-----------|-----------|-----------|
| $m=0$ | 1.812830 | 1.837006 | 1.854225 | 1.893239 | 1.911352 |
| $m=1$ | **1.814162** | **1.838208** | 1.857309 | 1.894418 | **1.912217** |
| $m=2$ | **1.814162** | **1.838208** | 1.856706 | 1.894442 | **1.912217** |
| $m=3$ | **1.814162** | **1.838208** | 1.856885 | 1.894444 | **1.912217** |
| $m=4$ | **1.814162** | **1.838208** | 1.856831 | **1.894445** | **1.912217** |
| $m=5$ | **1.814162** | **1.838208** | 1.856848 | **1.894445** | **1.912217** |
| $m=6$ | **1.814162** | **1.838208** | 1.856843 | **1.894445** | **1.912217** |
| $m=7$ | **1.814162** | **1.838208** | 1.856844 | **1.894445** | **1.912217** |
| $m=8$ | **1.814162** | **1.838208** | 1.856844 | **1.894445** | **1.912217** |
| $m=9$ | **1.814162** | **1.838208** | 1.856844 | **1.894445** | **1.912217** |

**Table 2**

|       | $E = 3.5$ | $E = 3.7$ | $E = 3.9$ | $E = 4.1$ | $E = 4.3$ | $E = 4.5$ |
|-------|----------|----------|----------|----------|----------|----------|
| $m=0$ | 0.197112 | 0.330442 | 0.882281 | 1.995810 | 0.727485 | 0.111258 |
| $m=1$ | 0.193031 | 0.324745 | 0.868192 | 1.957734 | 0.775876 | 0.123200 |
| $m=2$ | **0.193032** | **0.324751** | 0.868351 | 1.963232 | 0.778018 | 0.123285 |
| $m=3$ | **0.193032** | **0.324751** | **0.868347** | 1.962552 | 0.778127 | **0.123286** |
| $m=4$ | **0.193032** | **0.324751** | **0.868347** | 1.962637 | **0.778133** | **0.123286** |
| $m=5$ | **0.193032** | **0.324751** | **0.868347** | 1.962626 | **0.778133** | **0.123286** |
| $m=6$ | **0.193032** | **0.324751** | **0.868347** | 1.962628 | **0.778133** | **0.123286** |
| $m=7$ | **0.193032** | **0.324751** | **0.868347** | 1.962627 | **0.778133** | **0.123286** |
| $m=8$ | **0.193032** | **0.324751** | **0.868347** | 1.962627 | **0.778133** | **0.123286** |
| $m=9$ | **0.193032** | **0.324751** | **0.868347** | 1.962627 | **0.778133** | **0.123286** |

**Table 3**

|       | $E=1.0$ | $E=2.0$ | $E=3.0$ | $E=4.0$ | $E=5.0$ | $E=6.0$ | $E=7.0$ |
|---|---|---|---|---|---|---|---|
| $m=0$  | 1.103944 | 1.964438 | 0.487767 | 1.986578 | 1.301462 | 0.578798 | 0.046740 |
| $m=1$  | 1.145865 | 1.944708 | 0.201553 | 1.999994 | 1.409561 | 0.696655 | 0.047974 |
| $m=2$  | 1.145531 | 1.944649 | 0.239655 | 1.999993 | 1.410217 | 0.695754 | 0.048180 |
| $m=3$  | **1.145541** | **1.944628** | 0.266925 | **1.999996** | 1.410492 | 0.695974 | 0.048301 |
| $m=4$  | **1.145541** | **1.944628** | 0.266479 | **1.999996** | 1.410497 | **0.695971** | **0.048302** |
| $m=5$  | **1.145541** | **1.944628** | 0.267810 | **1.999996** | **1.410498** | **0.695971** | **0.048302** |
| $m=6$  | **1.145541** | **1.944628** | 0.267689 | **1.999996** | **1.410498** | **0.695971** | **0.048302** |
| $m=7$  | **1.145541** | **1.944628** | 0.267761 | **1.999996** | **1.410498** | **0.695971** | **0.048302** |
| $m=8$  | **1.145541** | **1.944628** | 0.267750 | **1.999996** | **1.410498** | **0.695971** | **0.048302** |
| $m=9$  | **1.145541** | **1.944628** | 0.267754 | **1.999996** | **1.410498** | **0.695971** | **0.048302** |
| $m=10$ | **1.145541** | **1.944628** | **0.267753** | **1.999996** | **1.410498** | **0.695971** | **0.048302** |
| $m=11$ | **1.145541** | **1.944628** | **0.267753** | **1.999996** | **1.410498** | **0.695971** | **0.048302** |
| $m=12$ | **1.145541** | **1.944628** | **0.267753** | **1.999996** | **1.410498** | **0.695971** | **0.048302** |

**Table 4**

|  | $E=1.0$ | $E=2.0$ | $E=3.0$ | $E=4.0$ | $E=5.0$ | $E=6.0$ | $E=7.0$ |
|---|---|---|---|---|---|---|---|
| $m=0$ | 1.103576 | 1.964775 | 0.427383 | 1.989915 | 1.305265 | 0.575136 | 0.048861 |
| $m=1$ | 1.120724 | 1.951618 | 1.471735 | 1.941171 | 1.578675 | 0.925720 | 0.168012 |
| $m=2$ | 1.120631 | 1.951608 | 0.366275 | 1.962159 | 1.594827 | 0.899100 | 0.170249 |
| $m=3$ | **1.120633** | 1.951590 | 1.808418 | 1.940044 | 1.605053 | 0.911032 | 0.174868 |
| $m=4$ | **1.120633** | **1.951591** | 0.158558 | 1.949353 | 1.605928 | 0.909080 | 0.174914 |
| $m=5$ | **1.120633** | **1.951591** | 1.692002 | 1.943866 | 1.606462 | 0.909649 | 0.175032 |
| $m=6$ | **1.120633** | **1.951591** | 0.019303 | 1.946573 | 1.606510 | 0.909533 | 0.175033 |
| $m=7$ | **1.120633** | **1.951591** | 1.747729 | 1.945127 | 1.606538 | 0.909562 | **0.175036** |
| $m=8$ | **1.120633** | **1.951591** | 0.098010 | 1.945870 | 1.606541 | 0.909556 | **0.175036** |
| $m=9$ | **1.120633** | **1.951591** | 1.720923 | 1.945481 | **1.606543** | 0.909557 | **0.175036** |
| $m=10$ | **1.120633** | **1.951591** | 0.061985 | 1.945683 | **1.606543** | 0.909557 | **0.175036** |
| $m=11$ | **1.120633** | **1.951591** | 1.734174 | 1.945578 | **1.606543** | 0.909557 | **0.175036** |
| $m=12$ | **1.120633** | **1.951591** | 0.080191 | 1.945632 | **1.606543** | 0.909557 | **0.175036** |